\documentclass[letterpaper,11pt]{article}
\usepackage{bbm}
\usepackage{latexsym,amssymb,amsmath, bm}

\usepackage{amsthm}
\usepackage{lipsum,graphicx,multicol}
\usepackage{threeparttable}
\usepackage{tabularx}

\usepackage[colorlinks=true,linkcolor = red,citecolor = blue]{hyperref}
\usepackage{setspace,color}
\usepackage[sort&compress]{natbib}
\usepackage{graphicx}
\usepackage[table]{xcolor}
\usepackage{float}
\usepackage{authblk}
\usepackage{leftidx}
\usepackage{geometry}
\usepackage[title,titletoc]{appendix}
\usepackage{epstopdf}
\usepackage{actuarialangle}
\usepackage{scrextend}
\usepackage{booktabs}
\usepackage{lscape}
\usepackage{caption}
\usepackage{subcaption}
\usepackage{indentfirst}
\usepackage{adjustbox}
\usepackage[linesnumbered,lined,boxed,commentsnumbered]{algorithm2e}
\allowdisplaybreaks

\numberwithin{equation}{section}
\numberwithin{figure}{section}
\numberwithin{table}{section}

\usepackage{enumitem}
\setlist{
  listparindent=\parindent,
  parsep=0pt,
}
\usepackage[title]{appendix}
\usepackage{multirow,bigdelim}
\usepackage{bbm}
\usepackage{tikz}
\usepackage{bm}

\usepackage{url}
\usepackage{breakurl}

\usetikzlibrary{decorations.pathmorphing}

\DeclareFontShape{OMX}{cmex}{m}{b}{<-> cmexb10}{}
\SetSymbolFont{largesymbols}{bold}{OMX}{cmex}{m}{b}
\usepackage{tcolorbox}
\usepackage{amsmath,bm}
\usepackage{arydshln}
\usepackage{mathtools}
\usepackage{mhchem}
\usepackage{gensymb}

\usepackage{accents}


\usepackage[table]{xcolor}

\definecolor{darkread}{rgb}{0.7, 0, 0}
\definecolor{darkbrown}{rgb}{0.55, 0.2, 0.15}
\definecolor{darkblue}{rgb}{0.1,0.1,0.6}
\definecolor{darkgreen}{rgb}{0.1,0.5,0.2}

\begin{document}

\doublespace
\title{\Large\textbf{Unveiling Nonlinear Dynamics in Catastrophe Bond Pricing: A Machine Learning Perspective}\thanks{We extend our gratitude for the valuable feedback received during various academic events, including the 2023 China International Conference on Insurance and Risk Management, the Actuarial Research Conference 2023, the 2023 INFORMS Annual Meeting, as well as from the engaging seminars held at the University of Connecticut, University of Waterloo, University of Science and Technology of China, and Nankai University. The usual disclaimer applies. Hong Li is supported in part by funding from the Social Sciences and Humanities Research Council [430-2022-00401].}}

\author{\qquad {Xiaowei Chen}\thanks{School of Finance, Nankai University, Tianjin 300350, China. Email: \href{mailto:chenx@nankai.edu.cn}{\color{magenta}chenx@nankai.edu.cn}}
\qquad {Hong Li}\thanks{Department of Economics and Finance, Gordon S. Lang School of Business and Economics, University of Guelph, Guelph, Canada. Email: \href{mailto:lihong@uoguelph.ca}{\color{magenta}lihong@uoguelph.ca} }
\qquad {Yufan Lu}\thanks{Department of Economics, The University of Melbourne, Melbourne, VIC 3010, Australia. Email: \href{yufanlu0121@gmail.com}{\color{magenta}yufanlu0121@gmail.com}}
\qquad {Rui Zhou}\thanks{Department of Economics, The University of Melbourne, Melbourne, VIC 3010, Australia. Email: \href{mailto:rui.zhou@unimelb.edu.au}{\color{magenta}rui.zhou@unimelb.edu.au}}}

\date{\today}


\maketitle

\begin{abstract}


This paper explores the implications of using machine learning models in the pricing of catastrophe (CAT) bonds. By integrating advanced machine learning techniques, our approach uncovers nonlinear relationships and complex interactions between key risk factors and CAT bond spreads---dynamics that are often overlooked by traditional linear regression models. Using primary market CAT bond transaction records between January 1999 and March 2021, our findings demonstrate that machine learning models not only enhance the accuracy of CAT bond pricing but also provide a deeper understanding of how various risk factors interact and influence bond prices in a nonlinear way. These findings suggest that investors and issuers can benefit from incorporating machine learning to better capture the intricate interplay between risk factors when pricing CAT bonds. The results also highlight the potential for machine learning models to refine our understanding of asset pricing in markets characterized by complex risk structures.

\end{abstract}
\noindent
{{\bf Keywords and phrases}: Catastrophe Bond, Asset Pricing, Machine Learning, Nonlinear Relationships, Conformal Prediction}


\noindent
{{\bf JEL Classifications}:} C22, C51, G11



\section{Introduction}


Catastrophe (CAT) bonds play a crucial role in transferring and managing the financial risks associated with natural disasters. These instruments provide a valuable source of capital for issuers, helping them mitigate potentially devastating financial losses, while offering investors an opportunity to diversify their portfolios. Understanding how CAT bond prices are determined hence is essential for both investors, who seek to make informed decisions, and issuers, who aim to optimize their risk management strategies. However, accurately pricing CAT bonds is a challenging task, as prices are influenced by a wide array of factors, including bond-specific characteristics and prevailing market conditions. This paper adopts a machine-learning-based approach to explore the complex relationships driving CAT bond prices, with a focus on uncovering potential nonlinear impacts and interaction effects among key risk factors. By shedding light on these intricate dynamics, the study contributes to a deeper understanding of asset pricing in the context of CAT bonds.

Early research by \cite{lane2000pricing}, which introduced a log-linear regression model, laid the foundation for the CAT bond pricing literature. Lane identified two key determinants of CAT bond prices in the primary market: the probability of first loss---the likelihood that the bond will experience any loss of principal due to a triggering event---and the conditional expected loss, which is the average percentage of the bond principal expected to be lost if the bond is triggered. Subsequent studies built on this linear framework by incorporating additional predictors and exploring pricing under various conditions. For instance, \cite{gurtler2016impact} included bond characteristics such as trigger type and bond rating, while \cite{braun2016pricing} integrated market condition indices including the Lane Synthetic Rate on Line index and the BB corporate bond spread. \cite{gotze2020hard} investigated sponsor-related pricing inefficiencies across different market conditions, and \cite{morana2019climate} examined the impact of climate change on CAT bond returns. Expanding the research scope further, \cite{braun2022common} developed factor pricing models for cross-sectional CAT bond returns, and \cite{herrmann2023trading} analyzed liquidity premiums in the secondary market. \cite{chatoro2023catastrophe} also explored issuer effects in the primary market. However, relying solely on linear models can oversimplify CAT bond price dynamics by overlooking nonlinear relationships and interactions between predictors.

In addition to using risk factors to predict the price of CAT risks, several other approaches have been proposed.  For instance, \cite{lee2002pricing} examined the pricing of default-risky CAT bonds in the context of moral hazard and basis risk; \cite{niehaus2002allocation} explored the efficiency (or lack thereof) in the allocation of catastrophe risk among individuals, primary insurers, and reinsurers; \cite{duan2005fair} analyzed how catastrophe risk affects the fair premiums for insurance guaranty funds; \cite{froot2008pricing} investigated the role of intermediated risks in catastrophe reinsurance; \cite{chang2010pricing} priced an Asian CAT option using a non-traded underlying loss index; \cite{lee2007valuation} introduced a contingent-claim framework for valuing catastrophe reinsurance contracts and CAT bonds; \cite{perrakis2013valuing} proposed a stochastic dominance approach to price catastrophe derivatives under limited diversification; \cite{braun2019asset} examined the determining factors affecting the returns of funds specializing in insurance-linked securities; \cite{zhao2020predicting} used actual catastrophe data to forecast CAT bond prices through market-based methods; and \cite{beer2022market} estimated smooth intensity surfaces for catastrophe arrivals from secondary CAT bond data.

More recently, machine learning techniques have been widely adopted in the financial asset pricing literature due to their ability to capture complex patterns and relationships \citep[see, among many others,][]{dong2022anomalies,jiang2023re,kelly2024virtue}. This trend is also evident in CAT bond pricing, where machine learning models have been increasingly utilized to better capture the nonlinear dynamics and interactions between predictors. For example, \cite{makariou2021random} introduced the Random Forest method to predict CAT bond spreads, demonstrating superior predictive performance compared to traditional linear models. Additionally, \cite{gotze2020risk} and \cite{gotze2023forecasting} applied Random Forest and Neural Network techniques to enhance the goodness-of-fit of CAT bond prices.


While machine learning methods could cope with pricing dynamics beyond linear relationships, they often focus on point forecasts. However, considering that predictions often deviate from the actual prices due to market complexity and dynamism, incorporating uncertainties surrounding points forecasts is crucial. Generating prediction intervals for asset prices is standard practice in financial markets. When issuing a CAT bond, the process typically involves an initial offering where the issuer provides tentative spreads to potential investors. Repricing can also happen during the issuing process based on investor feedback and market conditions.
Therefore, bond issuers should consider not just a single estimate but a credible range of potential final spreads, especially spreads in unfavorable scenarios, to realistically appraise the cost of risk transfer. For investors, understanding the possible range where the actual price could fall within is also vital in making informed investment decisions. This paper aims to combine the superior ability of machine learning approaches to discover complex pricing dynamics with the benefit of probabilistic forecasts in CAT bond pricing.


Our first methodological contribution introduces the use of Extreme Gradient Boosting (XGBoost) \citep{chen2016} for modeling CAT bond spreads. XGBoost, a robust machine learning tool, employs sequential boosting of weak learners (typically decision trees) to create a strong predictive model. 
While well-established in other domains, XGBoost has not been applied to CAT bond pricing. Our research explores its potential by comparing its prediction accuracy with that of the widely used linear regression pricing models and Random Forest. More importantly, we also employ various interpretation tools, including feature importance and accumulated local effects plots, to understand the nonlinear relationships and complex interactions between different risk factors and bond spreads. 

The second contribution of our study applies Conformal Prediction for constructing prediction intervals around point forecasts from machine learning models. Conformal Prediction intervals, studied by \cite{shafer2008tutorial}, \cite{Angelopoulos2021}, and \cite{fontana2023conformal}, are constructed from the residuals of the trained machine learning model and offer guaranteed coverage of the true response values with user-specified probability. Conformal Prediction is distribution-free and highly flexible, applicable to any prediction model, including Random Forest, Gradient Boosting algorithms, and various Neural Networks. It can even be applied to linear regression models, although this offers little value since producing prediction intervals is straightforward for linear regression models. 

Empirically, our study analyzes primary CAT bond market transaction records from January 1999 to March 2021, utilizing data from multiple data sources including trade notes provided by Lane Financial LLC, the Artemis Deal Directory, reports from rating agencies, and market research conducted by prominent industry participants such as Swiss Re, Munich Re, Aon Benfield, and Guy Carpenter. 
Our results highlight several advantages of the XGBoost model over traditional linear regression models. First, our approach reveals the nonlinear effects of several important risk factors on CAT bond spreads, including expected loss, probability of first loss, issue size, and reinsurance market cycles. We also observe significant interactions between these risk factors. For example, we find that the spread of bonds with higher expected loss values is more sensitive to reinsurance market conditions. In addition, the impact of issue size is negatively correlated with expected loss for bonds with high expected loss values. These nonlinear relationships and complex interactions are disregarded by linear regression models but naturally captured by the XGBoost model. Furthermore, XGBoost effectively handles risk factors with high correlations, such as expected loss and probability of first loss, while linear regression models often yield unstable estimates and interpretational challenges due to highly correlated risk factors. In linear regression, the usual practice involves removing one of the correlated risk factors, but this approach could inadvertently eliminate valuable information inherent in the excluded variable, leading to a less optimal model. XGBoost preserves the information they carry and enhances the model performance by allowing for highly correlated variables. Last but not least, XGBoost consistently achieves higher out-of-sample forecasting accuracy and generates more informative prediction intervals through Conformal Prediction. These prediction intervals are, on average, 8\% narrower than those from linear models, yet they maintain comparable coverage rates. In summary, this paper extends the existing literature on CAT bond pricing by introducing a probabilistic machine learning approach that provides insights on the nonlinear effects of various risk factors on CAT bond spread, as well as probabilistic forecasts with improved accuracy.


The structure of this paper is as follows: Section \ref{sec:data} provides an overview of the primary market CAT bond transaction data and present an exploratory analysis. Section \ref{sec:methodology} outlines the methodologies employed in our study, including linear regression, XGBoost, and the Conformal Prediction approach. Sections \ref{sec:lr} and \ref{sec:rf} present the empirical analysis results and their interpretations using linear regression models and machine learning models, respectively. Section \ref{sec:prob_analysis} constructs Conformal Prediction intervals, evaluates the prediction performance of the two modelling approaches, and further analyse conditional impacts of various risk factors. Finally, Section \ref{sec:conclusion} offers concluding remarks.

\section{Catastrophe Bond Data}
\label{sec:data}

The overall return on a CAT bond for the investor consists of two components. First, the investor receives a principal repayment at maturity. If the bond is not triggered, meaning no specified catastrophic events occur during its term, the full principal is repaid. However, if the bond is triggered by one or more specified events, the repayment amount is reduced.

Second, the investor receives coupon payments, which are pre-determined and remain unaffected by whether the bond is triggered. CAT bonds are typically issued as floating-rate securities, where the investor earns a fixed coupon spread over a reference index or the return on high-quality collateral, usually invested in short-term money market funds. The index or collateral return compensates investors for the time value of money and is independent of the bond’s embedded insurance risk. This rate resets periodically based on current short-term interest rates. The spread on a CAT bond compensates investors for assuming the insurance risk associated with potential catastrophic events.

The coupon rate is determined by the following formula:
\[
\text{Coupon Rate} = \text{Risk-Free Rate} + \text{CAT Bond Spread}
\]
Here, the \textit{Risk-Free Rate} refers to the return on the reference index, while the \textit{CAT Bond Spread} represents the additional yield that reflects the risk inherent in the bond. A higher CAT bond spread increases the coupon rate, thereby enhancing the overall return to the investor as compensation for the increased risk. Consequently, understanding the determinants of the CAT bond spread is essential for accurate pricing, a topic that has been extensively studied in the literature.

Our empirical study focuses on the primary market transactions of CAT bonds. A significant challenge in empirical research on CAT bond pricing is the limited availability of transaction data. To tackle this issue, we follow existing empirical studies by aggregating and verifying information from diverse sources. The dataset of CAT bonds used in this study encompasses a total of 765 bond tranches issued between January 1999 and March 2021. This dataset includes various critical bond characteristics, such as expected loss, probability of first loss, covered perils and territories, the trigger mechanism, sponsor information, and credit rating. The data was compiled from trade notes sourced from Lane Financial LLC\footnote{Source: \url{http://lanefinancialllc.com/}.}, the Artemis Deal Directory\footnote{Source: \url{https://www.artemis.bm/deal-directory/}.}, reports from credit rating agencies, and market research conducted by Swiss Re, Munich Re, Aon Benfield, and Guy Carpenter.

\begin{figure}[h!]
\begin{center}
		\begin{subfigure}[b]{0.5\textwidth}
			\includegraphics[width=\linewidth]{spread_EL_fullsample.pdf}
	           \caption{Expected loss}
	           \label{fig:spread_EL_sample}
		\end{subfigure}%
		\begin{subfigure}[b]{0.5\textwidth}
			\includegraphics[width=\linewidth]{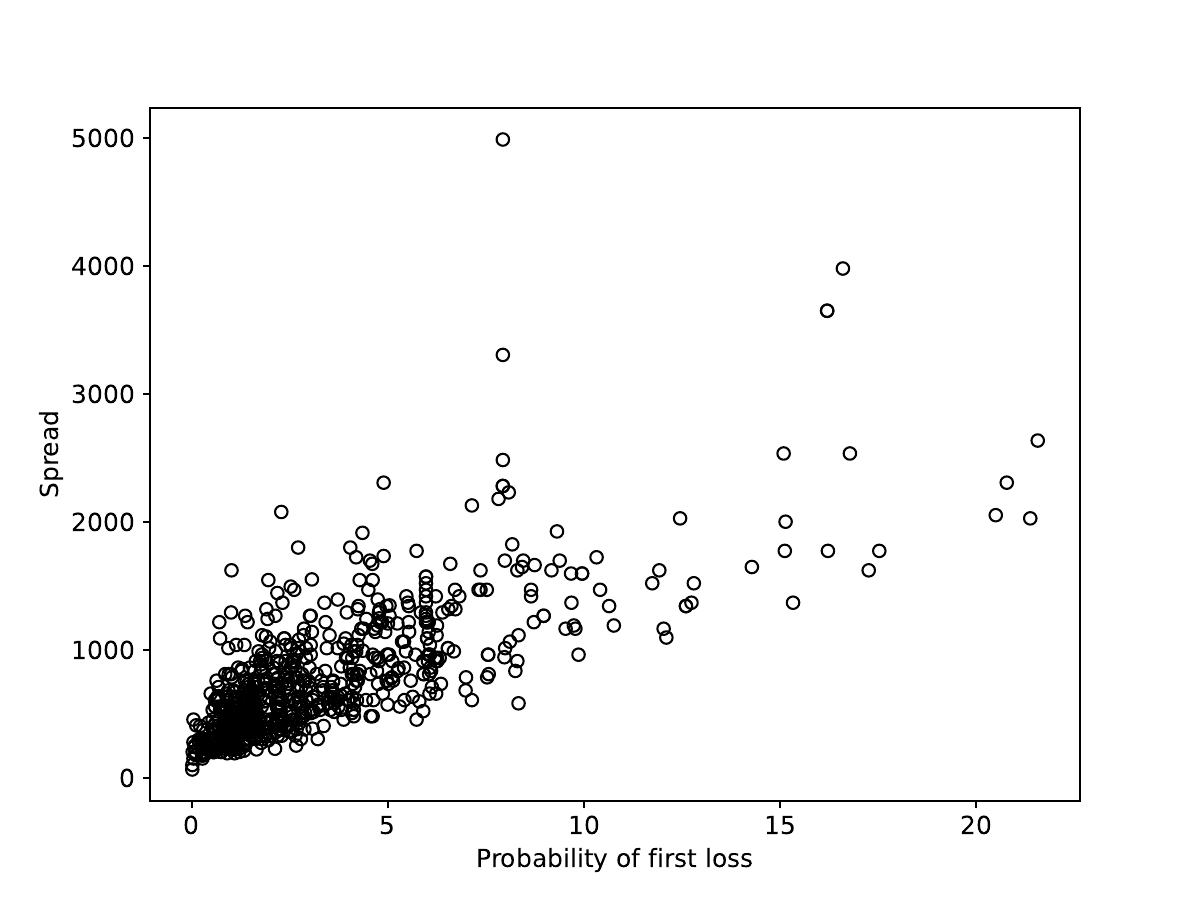}
	           \caption{Probability of first loss}
	           \label{fig:spread_PFL_sample}
		\end{subfigure}%
\end{center}
\caption{Scatter plots of bond spread against expected loss (left) and probability of first loss (right) using the 765 bonds in the data set. The horizontal axes are percentage points. }
\end{figure}

We begin with an exploratory analysis of the dataset. Figures~\ref{fig:spread_EL_sample} and \ref{fig:spread_PFL_sample} depict scatter plots of bond spread in relation to expected loss, the average percentage of the bond principal that is anticipated to be lost over the life of the bond due to triggering catastrophic events, and the probability of first loss. We can see a strong positive correlation between bond spread and both expected loss and probability of first loss in the sample. While the majority of bonds exhibit expected loss values below 5\% and probabilities of first loss below 10\%, there are some bonds with relatively high values of these factors. Furthermore, a handful of bonds demonstrate exceptionally large spreads relative to their expected loss or probability of first loss. This observation reveals that factors beyond expected loss and probability of first loss may affect the bond spread.

\begin{figure}[htbp]
\begin{center}
		\begin{subfigure}[b]{0.45\textwidth}
			\includegraphics[width=\linewidth]{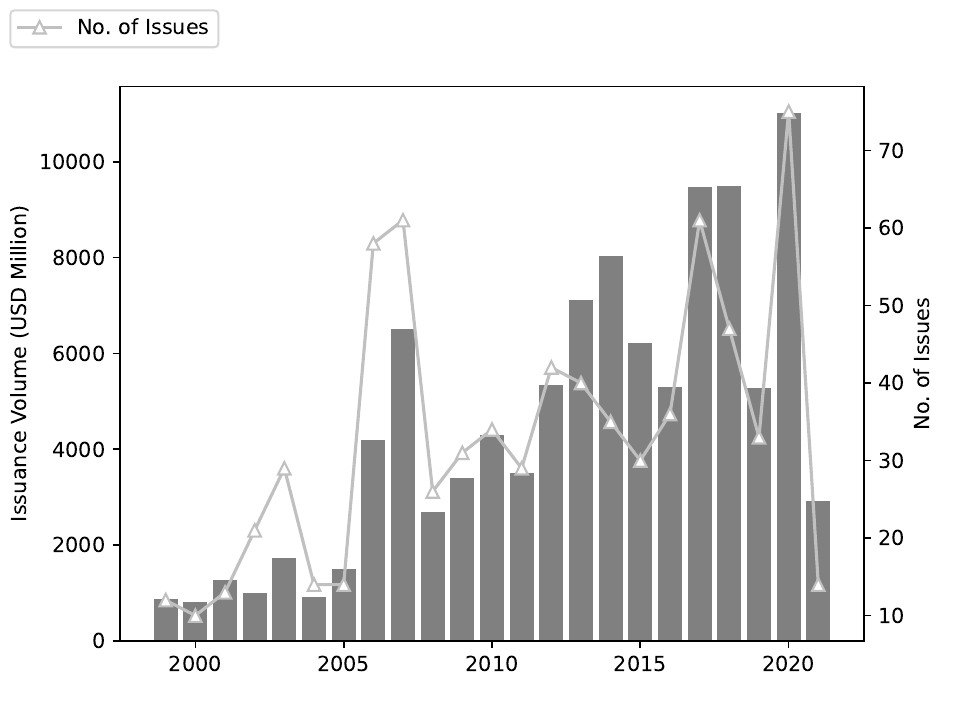}
	           \caption{Year}
	           \label{fig:f3a}
            \end{subfigure}%
		\begin{subfigure}[b]{0.45\textwidth}
			\includegraphics[width=\linewidth]{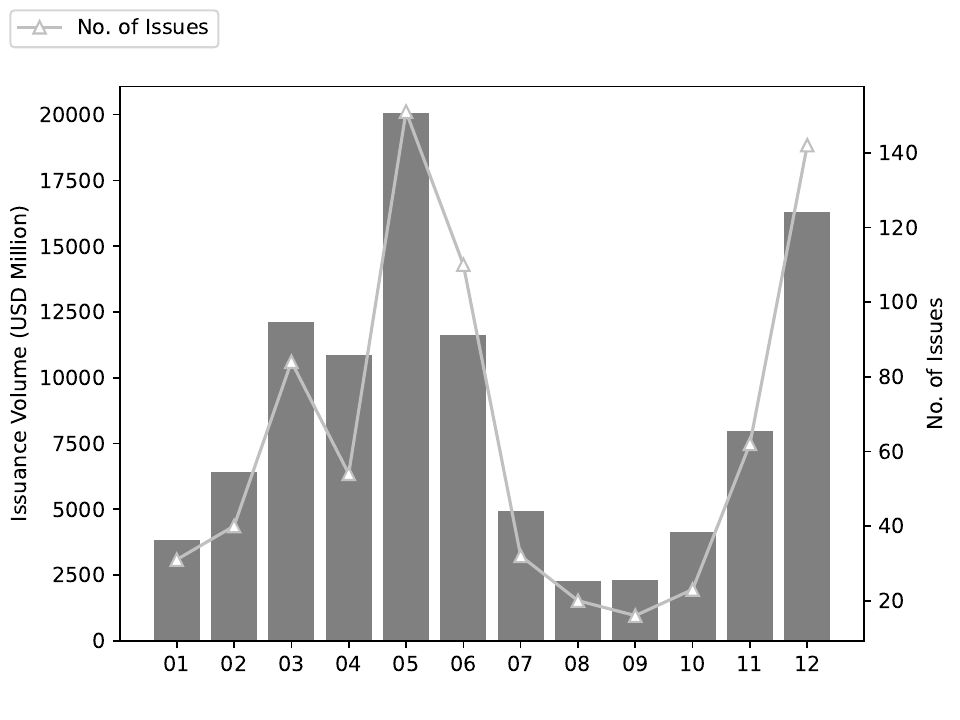}
	           \caption{Calendar month}
	           \label{fig:f3b}
		\end{subfigure}%
\end{center}
\caption{The number of CAT bonds issued and the issue volume per year (left) and per calendar month (right).}
\end{figure}

Figure \ref{fig:f3a} offers an overview of the volume and number of new transactions per year in the primary CAT bond market. We observe a sharp drop in CAT bond volume in 2008 during the financial crisis. This decline can be attributed to the flight-to-quality phenomenon, where investors tend to avoid risky assets during periods of market stress \citep{caballero2008collective}. 
However, investor confidence gradually rebounded, leading to a resurgence in CAT bond issuance in 2009 and 2010, with peak activity occurring in 2020. The surge in CAT bond issuance in 2020 was largely influenced by the COVID-19 pandemic, which motivated new issuers to enter the CAT bond market \citep{chatoro2023catastrophe}. Figure \ref{fig:f3b} presents the total issuance volume and the number of new transactions per calendar month. This chart demonstrates that the primary CAT bond market experiences heightened activity during the first, second, and fourth quarters of the year, while activity tends to be quieter in the third quarter. 
It is interesting to note that our findings differ from those of \cite{braun2016pricing}, which suggested relative inactivity in the first quarter. This discrepancy may arise from the inclusion of more recent transaction data in our study, which extends up to 2021, compared to the data used in Braun's analysis, which covers up to December 2012. Over the past decade, the frequency and severity of natural disasters have increased, potentially contributing to greater demand for coverage in the first quarter. The expansion of the insurance-linked securities market is another potential contributor to the increased activity in the first quarter.

In addition to bond-specific data, we consider several market-level indices to proxy the reinsurance market and interest rate cycles. We employ the Lane Financial LLC Synthetic Rate on Line Index ($\mathrm{ROLX}$), published quarterly by Thomson Reuters ILS Community, to represent the global reinsurance cycle. $\mathrm{ROLX}$ integrates data from the catastrophe bond, insurance-linked security (ILS), and industry-loss warranty (ILW) markets, thereby reflecting the average premiums paid for ILS and CAT bond-backed reinsurance or retrocession. In addition, we utilize the Guy Carpenter (GC) Global Property Catastrophe Rate on Line indices, covering different regions---Global, US, Asia Pacific, Europe, and UK---as indicators of reinsurance premium levels specific to catastrophe risk. These GC indices, published annually, monitor property catastrophe reinsurance rate-on-line movements in major global catastrophe reinsurance markets. They are based on reinsurance costs and reflect the expenses incurred in protecting property against catastrophic losses. The GC indices were previously employed by \cite{gurtler2016impact} as predictors of CAT bond spreads. \textcolor{black}{A larger value of these reinsurance market indices suggests an upswing in the reinsurance market cycle (i.e., a hard market), when premiums increase, coverage terms are restricted, and capacity for most types of reinsurance decreases.} Furthermore, we incorporate the Bank of America Merrill Lynch US High Yield BB Option-Adjusted Spread (BB spread), which is published daily, to capture the influence of the corporate bond markets. This index is calculated as the difference between a yield index for the BB rating category and the Treasury spot curve.

\begin{figure}[h!]
\begin{center}
		\begin{subfigure}[b]{0.5\textwidth}
			\includegraphics[width=\linewidth]{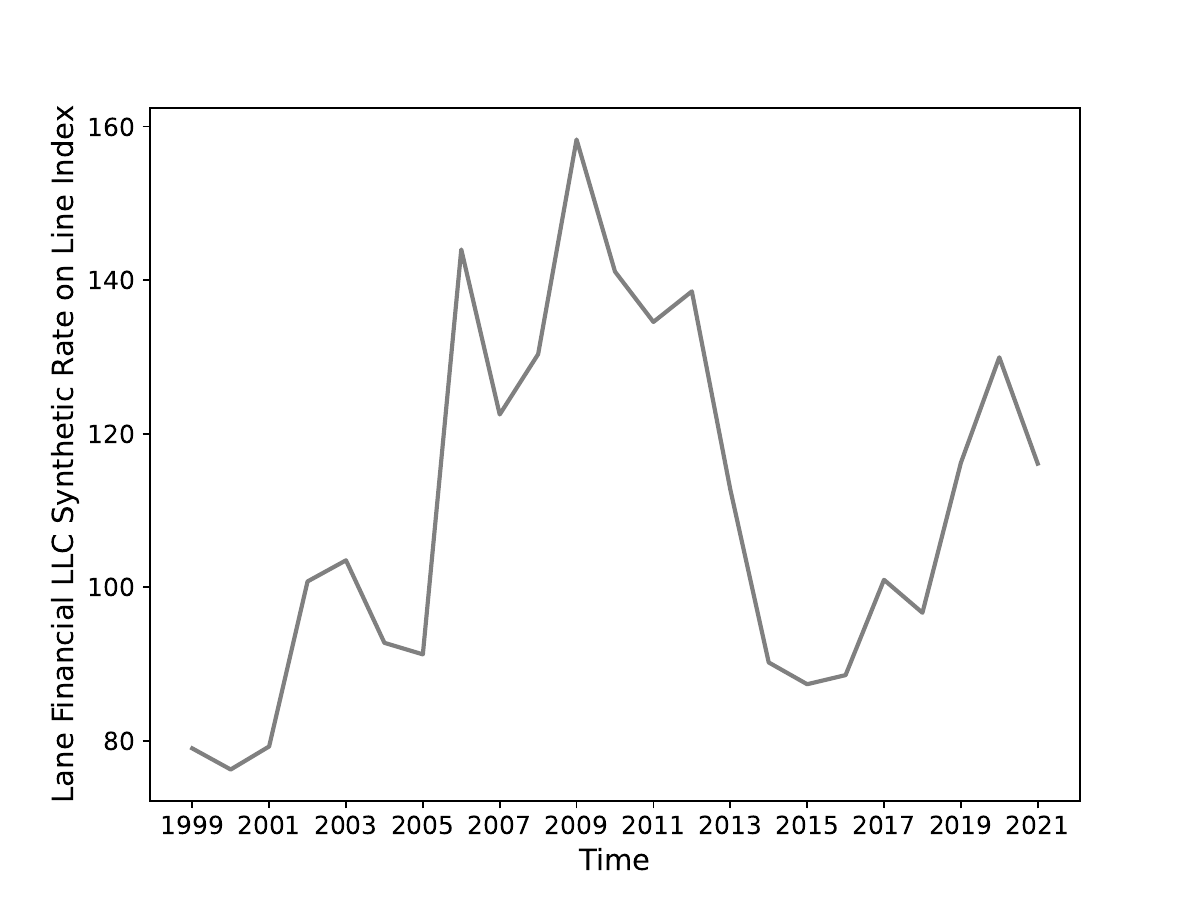}
	           \caption{Synthetic rate on line index}
	           \label{fig:Lane Financial LLC}
		\end{subfigure}%
		\begin{subfigure}[b]{0.5\textwidth}
			\includegraphics[width=\linewidth]{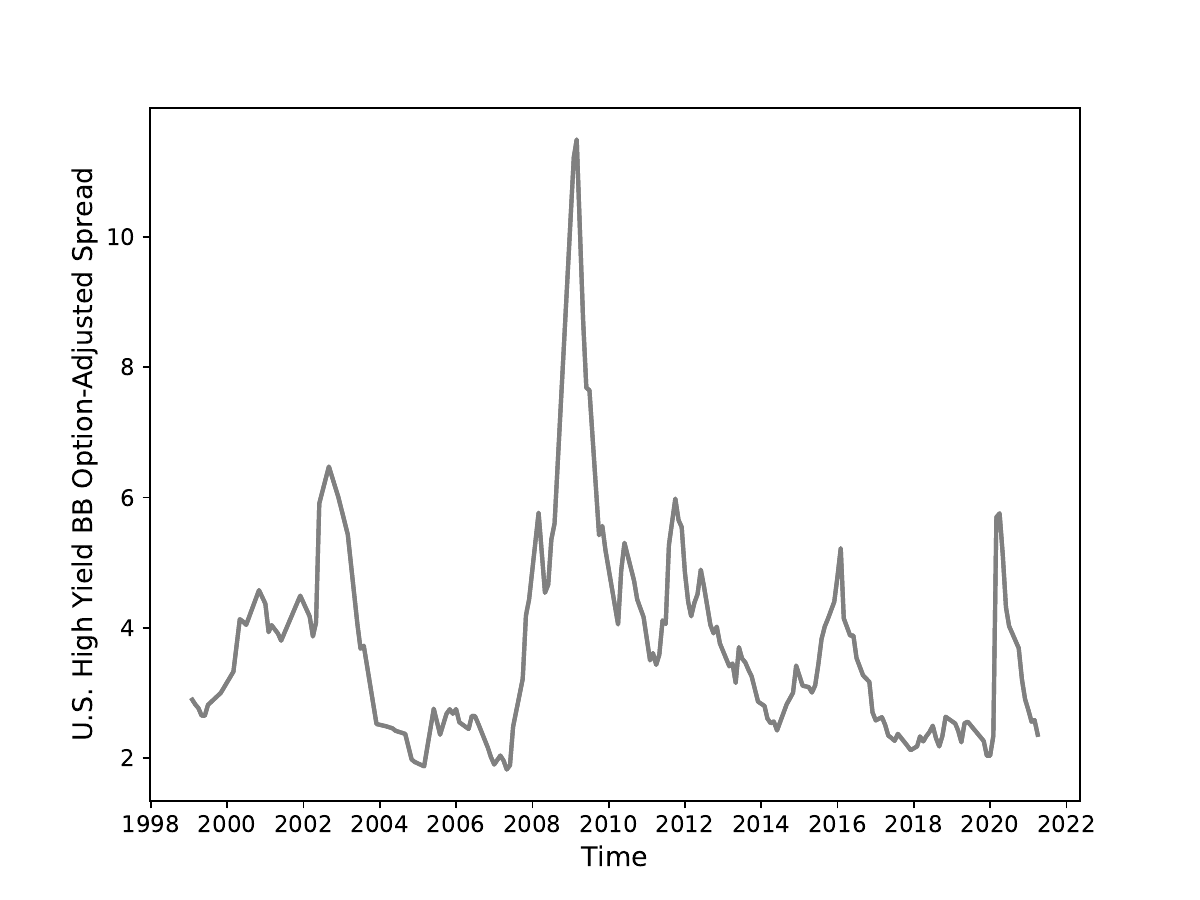}
	           \caption{US High Yield BB Option-Adjusted Spread}
	           \label{fig:U.S. High Yield BB}
		\end{subfigure}%
\end{center}
\caption{Lane Financial LLC Synthetic Rate on Line Index (left) and the US High Yield BB Option-Adjusted Spread (right) between January 1999 and March 2021.}
\label{fig:rolx_BB}
\end{figure}

\begin{figure}[h!]
\begin{center}
		\begin{subfigure}[b]{0.5\textwidth}
			\includegraphics[width=\linewidth]{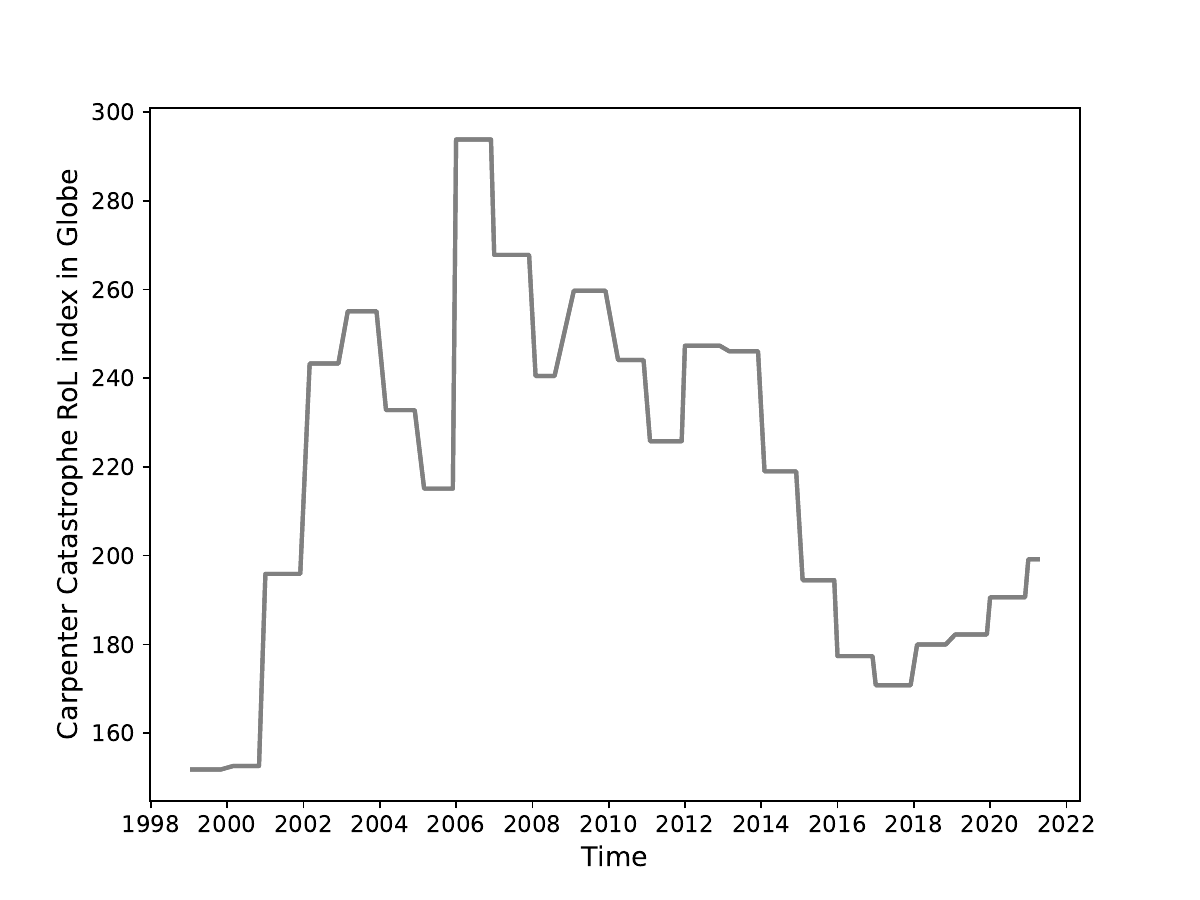}
	           \caption{Guy Carpenter Global RoL Index}
	           \label{fig:GC_glob}
		\end{subfigure}%
		\begin{subfigure}[b]{0.5\textwidth}
			\includegraphics[width=\linewidth]{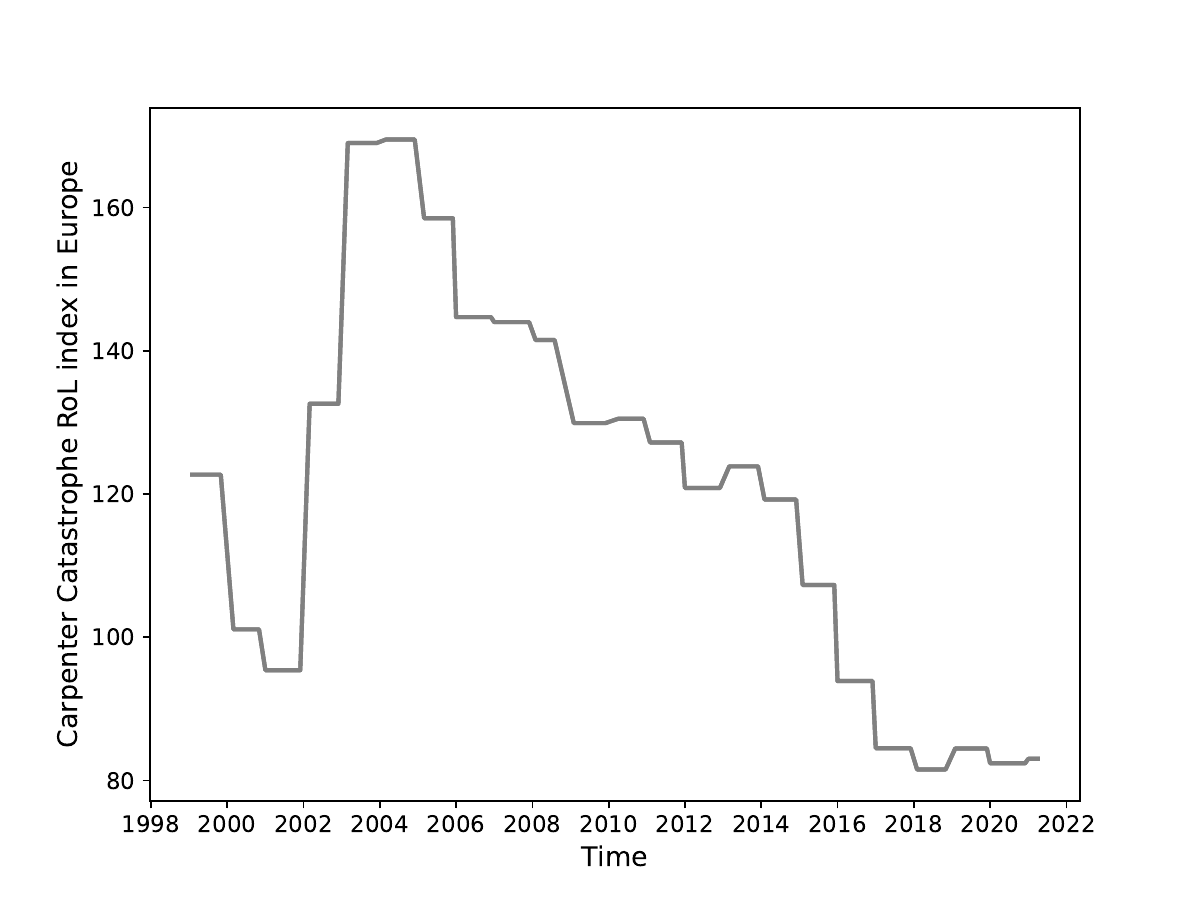}
	           \caption{Guy Carpenter European RoL Index}
	           \label{fig:GC_EU}
		\end{subfigure}\\
  		\begin{subfigure}[b]{0.5\textwidth}
			\includegraphics[width=\linewidth]{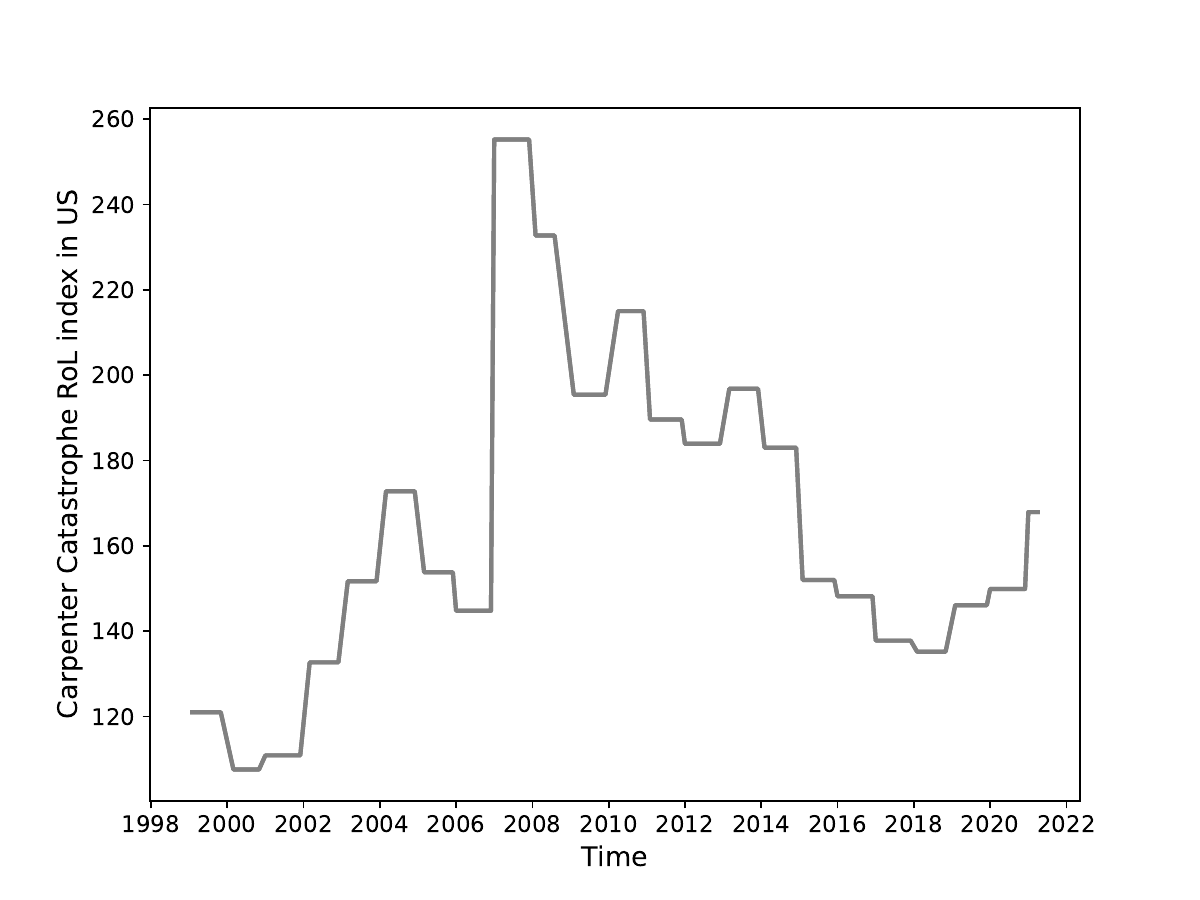}
	           \caption{Guy Carpenter US RoL Index}
	           \label{fig:GC_USA}
		\end{subfigure}%
		\begin{subfigure}[b]{0.5\textwidth}
			\includegraphics[width=\linewidth]{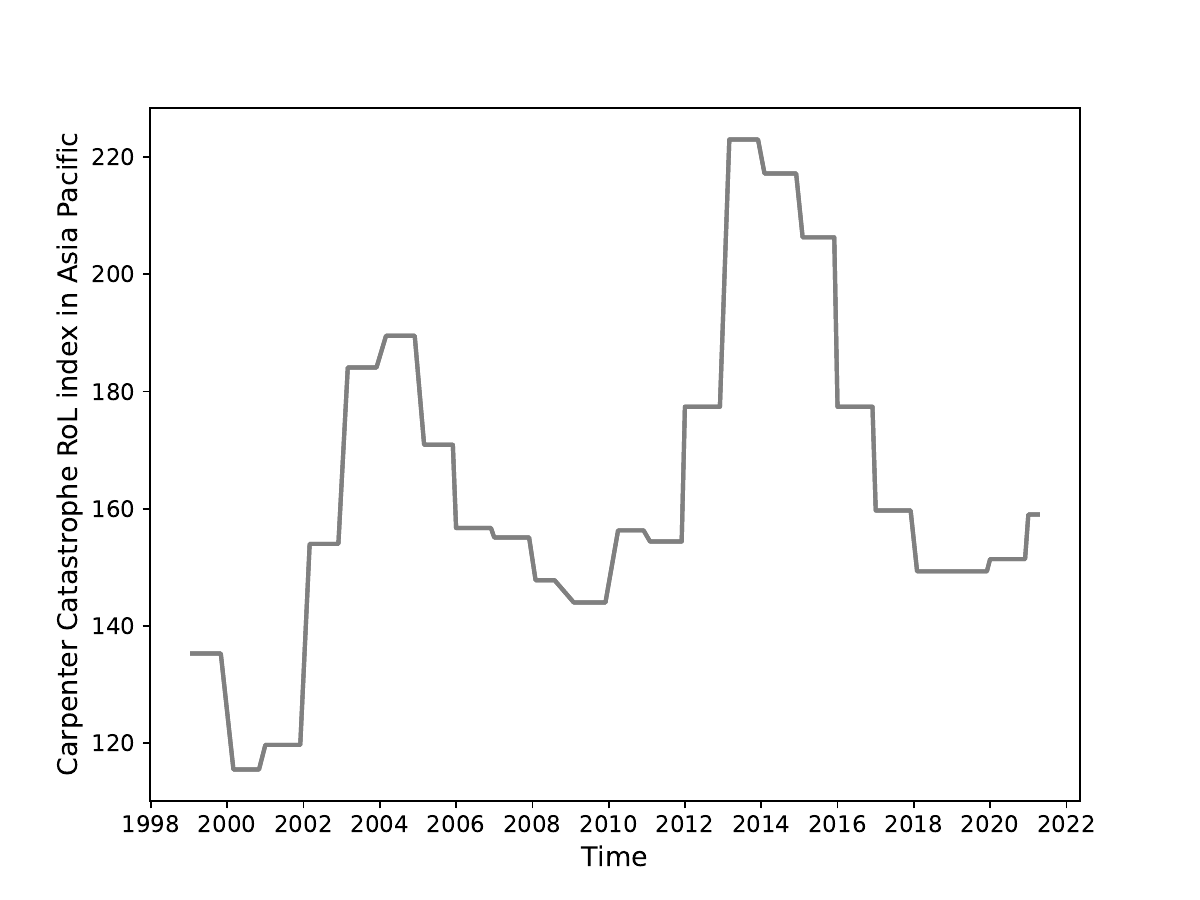}
	           \caption{Guy Carpenter Asia Pacific RoL Index}
	           \label{fig:GC_AP}
		\end{subfigure}%
\end{center}
\caption{The Guy Carpenter Rate on Line Index for the Global, European, US, and Asia Pacific markets, respectively, between January 1999 and March 2021.}
\label{fig:GC_indices}
\end{figure}

Figure~\ref{fig:rolx_BB} depicts the historical trends of ROLX and BB spread from January 1999 to March 2021. We can see that both indices peaked in 2009 due to the financial crisis. In addition, we observe a rising trend in ROLX since 2015, which suggests that global reinsurance demand has been growing faster than the reinsurance capacity, and reinsurance premiums have been increasing over this period. Figure~\ref{fig:GC_indices} illustrates the Guy Carpenter index for the Global, European, US, and Asia Pacific markets. The GC index for UK closely mirrors the European index and is thus omitted from the visualization. While these indices share similar overall trends, they also display distinct patterns. For instance, the global index initially peaked in 2003, experienced a slight decrease, and then surged to a higher peak around 2007. The European index reached its peak in 2005 before gradually declining. The US index spiked in 2008, during the financial crisis. The Asia-Pacific market exhibited two peaks, in 2004 and 2014, respectively.

Tables \ref{Descriptive} and \ref{Different Categorie} present a range of descriptive statistics for our sample of 765 CAT bonds. Table \ref{Descriptive} displays the mean, median, standard deviation, maximum, and minimum values for some key variables in the sample. The mean spread across all bonds is 771.38 bps, and the mean expected loss is 244.01 bps. The average risk premium per bond is 527.36 bps, calculated as the difference between spread and expected loss.  The spread exhibits considerable variability with values ranging from 66 bps to 4,988 bps. The highest spread is reported by Successor Hurricane Industry Ltd. (Series 3), which was identified as an outlier in \cite{braun2016pricing}. However, since our analysis focuses on machine learning methods that are more robust to outliers than linear regression models, we choose to retain this record in our sample. Similarly, expected losses vary widely, spanning from 1 bps to 1,484 bps. The sample includes CAT bond tranches with sizes ranging from USD \$1.80 million to USD \$1.5 billion, with a median size of USD \$100 million. The CAT bond tranche with an exceptionally large size of \$1.5 billion is from Everglades Re 2014-1 A which covers Florida hurricanes. Terms for CAT bonds in the sample vary from 5 months to 73 months, with a median maturity period of 37 months. 

\begin{table}[htbp]
 \centering
  \begin{threeparttable}
    \caption{Descriptive statistics for the 765 CAT bonds in the sample.}
     \begin{tabular}{lrrrrrr}
    \toprule
{} &    Mean &   S.D. &   Min &   Median &     Max \\
    \midrule
Spread (bps) & 771.38 & 514.64 & 66.00 & 634.00 & 4988.00 \\
Expected Loss (bps) & 244.01 & 233.12 & 1.00 & 160.00 & 1484.00 \\
Risk Premium (bps) & 527.36 & 368.88 & 65.00 & 449.00 & 4363.00 \\
Multiple & 6.24 & 19.08 & 1.19 & 3.71 & 456.00 \\
Size (USD mn)& 134.48 & 114.42 & 1.80 & 100.00 & 1500.00 \\
Term (months) & 37.36 & 12.13 & 5.00 & 37.00 & 73.00 \\



 \bottomrule
    \end{tabular}
     \label{Descriptive}
  \end{threeparttable}
\end{table}

Table \ref{Different Categorie} provides additional descriptive statistics for different categories of CAT bonds in the sample. An analysis of geographic classifications reveals that the majority of the sample consists of US bonds, making up 59.35\% of the total. The second-largest category comprises bonds covering multi-territory, accounting for 26.54\% of the sample. The frequency of bonds referencing a single territory other than the United States is relatively low, with Japan, Europe, and \textit{Other} territories representing 6.41\%, 6.67\%, and 1.05\% of the sample, respectively. Japanese peril bonds exhibit the highest average spreads, expected losses, and risk premiums, while bonds referencing \textit{Other} territory generally offer the lowest, likely due to their rarity and valuable diversification characteristics. The risk-return trade-off, measured by multiples of spread to expected loss, varies significantly across different territories. Multiterritory bonds offer the highest multiples at 7.21, while bonds categorized under \textit{Other} territories present the lowest multiples, at 2.65. Furthermore, Japanese bonds and bonds referencing \textit{Other} territory have the smallest and largest average sizes, respectively.

\begin{table}[htbp]
 \centering
  \begin{threeparttable}
    \caption{Descriptive statistics for different categories of CAT bonds in the sample.
}
     \begin{tabular}{p{3.5cm}p{1cm}p{1.5cm}p{1.5cm}p{1.2cm}p{1.2cm}p{1.5cm}p{1.8cm}p{1.5cm}}
\toprule
             &  No. &  Percent &   S\textsuperscript{CAT} (bps)  & EL  (bps) & RP (bps) & Multi (bps) & \centering Size (USD mn) & Term (months) \\
\midrule
\textbf{Territory} \\
Multiterritory & 203 & 26.54 & 788.12 & 276.38 & 511.74 & 7.21 & 113.93 & 39.49 \\
US & 454 & 59.35 & 753.31 & 215.88 & 537.43 & 6.31 & 149.50 & 36.85 \\
Europe & 51 & 6.67 & 513.67 & 193.47 & 320.20 & 4.86 & 126.25 & 38.69 \\
Japan & 49 & 6.41 & 1199.55 & 427.61 & 771.94 & 3.58 & 77.90 & 32.57 \\
Other & 8 & 1.05 & 392.00 & 217.12 & 174.88 & 2.65 & 202.50 & 32.88 \\
        \hline
Total & 765 & 100 & & & & & &\\
\\
\textbf{Peril} \\
Multiperil & 464 & 60.65 & 834.12 & 276.05 & 558.07 & 6.82 & 118.09 & 39.22 \\
Wind & 165 & 21.57 & 806.00 & 220.58 & 585.42 & 4.97 & 143.21 & 32.65 \\
Earthquake & 121 & 15.82 & 447.97 & 139.76 & 308.21 & 5.78 & 181.72 & 35.89 \\
Other & 15 & 1.96 & 1058.40 & 351.67 & 706.73 & 6.18 & 164.31 & 43.33 \\
       \hline
Total & 765 & 100 & & & & & &\\
\\
\textbf{Trigger} \\
Indemnity & 327 & 42.75 & 677.62 & 219.06 & 458.55 & 7.63 & 158.55 & 38.94 \\
Other & 438 & 57.25 & 841.37 & 262.64 & 578.73 & 5.20 & 116.50 & 36.17 \\
        \hline
Total & 765& 100 & & & & & &\\
\\
\textbf{Sponsor} \\
Other & 512 & 66.93 & 723.96 & 227.12 & 496.84 & 6.04 & 152.74 & 40.34 \\
Swiss Re & 253 & 33.07 & 867.34 & 278.21 & 589.13 & 6.65 & 97.52 & 31.32 \\
       \hline
Total & 765 & 100 & & & & & &\\

\bottomrule
    \end{tabular}
     \label{Different Categorie}
  \end{threeparttable}
\end{table}

Peril-specific figures show that wind, earthquake, and multiperil bonds account for 21.57\%, 15.82\%, and 60.65\% of the sample, respectively. Earthquake bonds demonstrate by far the lowest average spreads, expected losses, and risk premiums among all categories while bonds covering \textit{Other} peril have the highest. Among all peril categories, earthquake bonds have the largest average issue and \textit{Other} peril bonds exhibit the longest average term. Regarding trigger types, bonds with an indemnity trigger account for about 42.75\% of the sample. CAT bonds with indemnity triggers offer lower average spreads and risk premiums than those with other triggers, despite compensating investors for the moral hazard associated with the insured. This is due to indemnity-trigger bonds having on average lower expected losses. The risk-return trade-off appears attractive for indemnity trigger transactions, as they offer higher average multiples of spread to expected loss than transactions with other triggers by a significant margin. Finally, sponsors are distinguished between Swiss Re and all other institutions. Swiss Re accounts for nearly a third (33.07\%) of the tranches in our sample. Swiss Re deals feature higher average spreads and expected losses than other bonds, with their average multiples also being greater.

\section{Methodology} \label{sec:methodology}

This section introduces the statistical learning methods used in our analysis. Section \ref{sec:prob_formulation} formulates the CAT bond pricing problem, while Sections \ref{sec:lr_method} and \ref{sec:ensemble} introduce the linear model and ensemble machine learning methods, respectively. Finally, Section \ref{sec:conformal_prediction} outlines the Conformal Prediction approach.

\subsection{Problem Formulation}
\label{sec:prob_formulation}

Let us denote the spread of a CAT bond at issuance as $Y$, and let $\bm{X}=(X_1,X_2,\ldots,X_p)$ represent $p$ distinct risk factors. This vector $\bm{X}$ encompasses variables that convey information about bond characteristics, market conditions, and economic indicators. We assume there exists a relationship between $Y$ and $\bm{X}$, expressed as:
\[Y = f(\bm{X}) + \epsilon.\]
In this equation, $f$ is a fixed but unknown function of $X_1,\ldots, X_p$, and $\epsilon$ represents a random error term, independent of $\bm{X}$, with a mean value of zero. The function $f$ incorporates the systematic information provided by $\bm{X}$ regarding $Y$.

Assume that we have observed a collection of $n$ distinct data points from primary CAT bond market transactions. We use $x_{ij}$ to represent the value of the $j$th risk factor from the $i$th observation, where $i = 1, 2, \ldots ,n$ and $j = 1, 2, \ldots ,p$. Correspondingly, $y_i$ denotes the value of the final spread from the $i$th transaction. We can represent the observed dataset as $D=\{(\bm{x}_1, y_1), (\bm{x}_2, y_2), \ldots , (\bm{x}_n, y_n)\}$, where $\bm{x}_i = (x_{i1}, x_{i2},\ldots,x_{ip})^T$.

The estimation of $f$ involves exploring the hypothesis space $H$, which is the set of all feasible functions, to find a function, denoted as $\hat{f}$, that can capture the relationship between $\bm{X}$ and $y$ as accurately as possible. This process, commonly referred to as learning, is guided by minimizing a loss function that quantifies the difference between the true output $(y_1,y_2,\ldots,y_n)$ from the observed data and the predicted outputs $\bigl(\hat{f}(\bm{x}_1),\hat{f}(\bm{x}_2),\ldots,\hat{f}(\bm{x}_n)\bigr)$. 

\subsection{Linear Regression}
\label{sec:lr_method}

In the existing literature, linear regression is a popular choice for modeling (the transform of) CAT bond spreads. This method assumes a linear structure for $f$:
\[f(\bm{X}) = \beta_0 + \beta_1 X_1 + \beta_2 X_2 +\cdots + \beta_p X_p.\]
Typically, a linear regression model is estimated using ordinary least squares, where the estimates of the $p + 1$ coefficients, $\beta_0,\beta_1, \ldots,\beta_p$, minimize the sum of the squares of the residuals. Model estimation for linear regression does not require a specific distribution assumption for the error term $\epsilon$. However, for inference, which includes conducting statistical tests and calculating confidence and prediction intervals, a normal distribution assumption is commonly adopted.

\subsection{Ensemble Learning}
\label{sec:ensemble}

This subsection introduces XGBoost, an ensemble learning method used in our analysis. An overview of ensemble methods is provided in Section \ref{sec:overview}, while Section \ref{sec:gbm} explore the specifics of XGBoost. 

\subsubsection{Overview}
\label{sec:overview}

Ensemble learning is a widely adopted machine learning technique that leverages the power of multiple models (base learners) to collectively address a problem, resulting in improved accuracy and robustness compared to individual models. Instead of relying on a single function $f$ for estimation, ensemble learning combines the outputs of multiple functions that approximate $f$. In this study, we employ XGBoost, a well-established ensemble learning methods.

Ensemble learning uses base learners, often capable of capturing nonlinearity, such as decision trees, to model complex relationships between explanatory variables and the variable of interest. Consequently, ensemble learning frequently outperforms linear regression in model goodness-of-fit as well as prediction accuracy. However, the amalgamation of multiple models can make the output less interpretable. To better understand the model's behavior, various strategies, including partial dependence plots and feature importance measures, have been developed to illustrate the significance of individual explanatory variables in model predictions. 

\subsubsection{XGBoost}\label{sec:gbm}

XGBoost is a variation of the Gradient Boosting Machine (GBM), which was proposed by Friedman \citep{Friedman2001,FRIEDMAN2002}. GBM is an ensemble learning method that has gained popularity for its superior predictive performance and versatility in handling various variable types and loss functions. GBM builds an ensemble of weak models, typically decision trees, iteratively. In each iteration, GBM fits a new model to the residuals of the cumulative prediction of all previous models. This approach focuses more on data points that are challenging to predict accurately, improving overall performance. After training, these weak models are combined to form a strong learner capable of accurate predictions. The GBM estimator is represented as $\hat{f}(\bm{X})=\sum_{b=1}^{B}\eta\hat{f}_b(\bm{X})$, where $\hat{f}_b$ denotes the estimated model from the $b$th iteration, and $\eta$ is the learning rate that scales the contribution of each tree, aiming for controlling overfitting. A detailed description of the GBM algorithm is provided in Appendix \ref{apdx:gbm}, and for more detailed discussions on GBM, readers are referred to \cite{natekin2013gradient}.


XGBoost \citep{chen2016} extends GBM in several aspects.
Firstly, similar to Random Forest, XGBoost employs column sub-sampling, which reduces overfitting and enhances computational speed. Secondly, while standard GBM implementations stop splitting a node upon encountering negative loss, XGBoost grows the tree to its maximum depth and then prunes it backward to ensure optimal splits are not missed. 
Furthermore, XGBoost supports monotone and interaction constraints, which could improve model interpretability. Monotone constraints dictate that an increase in a particular feature should consistently lead to an increase (or decrease) in the predicted value of the response variable. This is particularly useful when prior knowledge suggests a monotonic relationship between a predictor and the response variable. For instance, all else being equal, a higher expected loss should lead to a higher bond spread. Interaction constraints specify which variables can interact with each other in the trees, making interactions more controllable and limiting model complexity. 

\subsection{Conformal Prediction}

\label{sec:conformal_prediction}

In this section, we introduce the Conformal Prediction algorithm, which we will use to generate prediction intervals for our ensemble methods. Section \ref{sec:prediction_region} defines the prediction interval, a concept fundamental to Conformal Prediction. Section \ref{sec:idea} provides an overview of Conformal Prediction, while Section \ref{sec:jackknife} introduces the \textit{Jackknife+} method, a popular choice of Conformal Prediction that we will employ in our analysis.

\subsubsection{Prediction Interval} \label{sec:prediction_region}

Ensemble methods like Random Forest and XGBoost only offer point predictions. Asset pricing and risk management problems typically require probabilistic predictions of the outcome. For instance,  we are often interested in the prediction uncertainty, such as the 95\% prediction region, which is an interval or set that contains the true value of the response variable with at least 95\% probability. 
Formally, consider a set of independent and identically distributed (i.i.d.) training data $\{(\bm{x}_1,y_1),\ldots,(\bm{x}_n,y_n)\}$ and a new observation $(\bm{x}_{n+1},y_{n+1})$ drawn independently from the same distribution. After fitting a model to the training data and obtaining the fitted model $\hat{f}$, we aim to construct a prediction interval around $\hat{f}(\bm{x}_{n+1})$ that contains the true test response value with at least $100(1-\alpha)\%$ probability. This prediction interval, denoted as $C_\alpha(\bm{x}_{n+1})$, is defined such that:
\begin{equation}\label{eq:PI}
    P\bigl(y_{n+1}\in C_\alpha(\bm{x}_{n+1})\bigr)\geq 1-\alpha.
\end{equation}

\subsubsection{Overview of Conformal Prediction} \label{sec:idea}

Conformal prediction is a versatile approach applicable to any pre-trained model for constructing prediction intervals that guarantee coverage of the true response value with a predefined probability \citep{fontana2023conformal,Angelopoulos2021,shafer2008tutorial}. Conformal Prediction is distribution-free, meaning that it could generate prediction intervals without any distributional assumptions of the error term. 
While Conformal Prediction is applicable to both classification and regression problems, we will illustrate how to construct a prediction interval with a regression problem since it is the focus of this paper.

The fundamental idea behind Conformal Prediction is related to a result for sample quantiles. To illustrate the idea, let us first consider a set of generic i.i.d. observations $u_1, \ldots, u_n$ from a scalar random variable. For a target coverage level $1-\alpha$, with $\alpha\in(0,1)$, and a new observation $u_{n+1}$ drawn independently from the same distribution of $u_1, \ldots, u_n$, we have:
\[P(u_{n+1}\leq\hat{q}_{1-\alpha}\{u_1,\ldots,u_n\})\geq 1-\alpha,\]
where $\hat{q}_{1-\alpha}\{u_1,\ldots,u_n\}$ represents the $(1-\alpha)$ sample quantile of $u_1, \ldots, u_n$. The sample quantile is calculated as:
\[
\hat{q}_{1-\alpha}\{u_1,\ldots,u_n\}=\left\{
\begin{array}{cc}
    u_{(\lceil (n+1)(1-\alpha)\rceil)} & \mbox{ if }\lceil (n+1)(1-\alpha)\rceil\leq n  \\
    \infty & \mbox{otherwise}, 
\end{array}
\right.
\]
where $u_{(1)}\leq \ldots \leq u_{(n)}$ denote the order statistics of $u_1, \ldots, u_n$, and $\lceil (n+1)(1-\alpha) \rceil$ is the smallest integer greater than or equal to $(n+1)(1-\alpha)$.

Based on this result, a naive way of constructing prediction intervals is to use residuals from the training data to estimate the typical prediction error for the new test data point. Returning to the training data $\{(\bm{x}_1,y_1),\ldots,(\bm{x}_n,y_n)\}$ and a new observation $(\bm{x}_{n+1},y_{n+1})$. Denoting the absolute residual for the $i$th data point as $r_i=|y_i-\hat{f}(\bm{x}_i)|$, the ``naive'' prediction interval can be constructed by adding and subtracting the $(1-\alpha)$ sample quantile of the absolute residuals from the point forecasts of $y_{n+1}$, $\hat{f}(\bm{x}_{n+1})$:
\begin{equation}
   C_\alpha(\bm{x}_{n+1})=\left[\hat{f}(\bm{x}_{n+1})-\hat{q}_{1-\alpha}\{r_1,\ldots,r_n\},\hat{f}(\bm{x}_{n+1})+\hat{q}_{1-\alpha}\{r_1,\ldots,r_n\}\right]. \label{eq:prediction_interval} 
\end{equation}
Although being rather straightforward, the interval in Equation~\eqref{eq:prediction_interval} may offer significantly lower coverage than the desired $1-\alpha$ level due to overfitting, as fitted residuals tend to be smaller than residuals on unseen test points. A number of methods have been proposed to address the underestimation issue of the naive prediction region. The subsection below will introduce a few of these methods, including the \textit{split-set}, the \textit{Jackknife}, and eventually the \textit{Jackknife+} method, which will be used in our analysis.


\subsubsection{Jackknife+ Conformal Prediction Interval} \label{sec:jackknife}

One way to avoid the overfitting issue of the naive method is to construct prediction intervals based on residuals from observations not seen during training. \cite{Papadopoulos2002InductiveCM} and \cite{Vovk2005} introduced the \textit{split-set} Conformal Prediction, which divides the training data into two subsets, $\mathcal{S}_1$ and $\mathcal{S}_2$. The model $\hat{f}$ is trained using $\mathcal{S}_1$, while absolute residuals are computed for $\mathcal{S}_2$ to inform the width of the prediction interval.

While splitting the training set reduces overfitting, it introduces extra randomness, as different splits can lead to different Conformal Prediction intervals. To mitigate this randomness, inferences from multiple splits can be combined to obtain a more stable interval. Here, we introduce the procedure of constructing a \textit{Jackknife} Conformal Prediction interval \citep{Barber2021}, which leverages sample quantiles of leave-one-out residuals for interval definition:
\begin{enumerate}
    \item Train the model using all observations in the training set, excluding the $i$th observation:\\
    $(\bm{x}_1,y_1),\ldots,(\bm{x}_{i-1},y_{i-1})$, $(\bm{x}_{i+1},y_{i+1}),\ldots,(\bm{x}_{n},y_{n})$, and denote it as $\hat{f}^{(-i)}$.
    \item Calculate the absolute residual for the $i$th observation, $r_i=|y_i-\hat{f}^{(-i)}(\bm{x}_i)|$, using $\hat{f}^{(-i)}$ and the predictor vector $\bm{x}_i$.
    \item Repeat Steps 1 and 2 for all $i=1,\ldots,n$.
    \item Determine the $(1-\alpha)$ sample quantile of the absolute residuals, $\hat{q}_{1-\alpha}\{r_1,\ldots,r_n\}$.
    \item Train the model using the entire training set $\{(\bm{x}_i,y_i), i=1\ldots,n\}$ and obtain the trained model $\hat{f}$.
    \item The \textit{Jackknife} Conformal Prediction interval for a new observation $(\bm{x}_{n+1},y_{n+1})$ is defined as:
    \[C_\alpha(\bm{x}_{n+1})=\left[\hat{f}(\bm{x}_{n+1})- \hat{q}_{1-\alpha}\{r_1,\ldots,r_n\},\hat{f}(\bm{x}_{n+1})+ \hat{q}_{1-\alpha}\{r_1,\ldots,r_n\}\right].\]
\end{enumerate}
While the \textit{Jackknife} method often empirically achieves $1-\alpha$ coverage, it does not guarantee finite sample coverage validity, indicating that, for a finite sample size, the interval may not contain the true value $100(1-\alpha)$\% of the time. Asymptotic validity is only ensured when non-trivial conditions are satisfied by the trained model \citep{steinberger2016}. The \textit{Jackknife} prediction interval may exhibit poor coverage when the trained model $\hat{f}$ is unstable due to its lack of finite sample coverage validity.

To address this issue, \cite{Barber2021} introduced the \textit{Jackknife+} method, which copes better with small data sets. 
It modifies the \textit{Jackknife} procedure by computing $\hat{f}^{(-i)}(\bm{x}_{n+1})$ for $i=1,\ldots,n$ and defines the \textit{Jackknife+} Conformal Prediction interval as the interval between the $\alpha$ sample quantile of $\bigl\{\hat{f}^{(-i)}(\bm{x}_{n+1})-r_i\bigr\}$, $i=1,\ldots,n$, and the $(1-\alpha)$ quantile of $\bigl\{\hat{f}^{(-i)}(\bm{x}_{n+1})+r_i\bigr\}$, $i=1,\ldots,n$:
\[C_\alpha(\bm{x}_{n+1})=\left[\hat{q}_\alpha \left\{\hat{f}^{(-i)}(\bm{x}_{n+1})-r_i,i=1,\ldots,n\right\} , \hat{q}_{1- \alpha}  \left\{\hat{f}^{(-i)}(\bm{x}_{n+1})+r_i,i=1,\ldots,n\right\} \right].\]
The \textit{Jackknife+} method is guaranteed to provide at least $1-2\alpha$ coverage and often empirically achieves $1-\alpha$ coverage, even for small datasets. In the subsequent empirical analysis, we will use the \textit{Jackknife+} method to generate prediction intervals of the ensemble learning methods.



\section{Linear Regression: Model Estimation and Implications}
\label{sec:lr}

Our analysis begins with constructing an extensive linear regression model, incorporating a range of risk factors considered in existing studies \citep{braun2016pricing,gurtler2016impact,beer2022market}. The list of risk factors and their descriptions are provided in Table \ref{tbl:glossary}. The linear regression model takes the following form:
\begin{equation}
\begin{aligned}
 S_i^{\mathrm{CAT}}= &\alpha+ \beta_{\mathrm{EL}} \mathrm{EL}_i  +\beta_{\mathrm{PFL}} \mathrm{PFL}_i +\beta_{\mathrm{SIZE}} \mathrm{SIZE}_i+\beta_{\mathrm{TERM}} \mathrm{TERM}_i+\beta_{\mathrm{INDEM}} \mathrm{INDEM}_i \\
 & +\beta_{\mathrm{WIND}} \mathrm{WIND}_i+\beta_{\mathrm{EQ}} \mathrm{EQ}_i+\beta_{\mathrm{US}} \mathrm{US}_i+\beta_{\mathrm{EU}} \mathrm{EU}_i+\beta_{\mathrm{JP}} \mathrm{JP}_i+\beta_{\mathrm{SR}} \mathrm{SR}_i  \\
&+\beta_{\mathrm{IG}}\mathrm{IG}_i+\beta_{\mathrm{ROLX}} \mathrm{ROLX}_i+\beta_{\mathrm{BBSPR}} \mathrm{BBSPR}_i +\beta_{\mathrm{GC.GLOB}} \mathrm{GC.GLOB}_i +\beta_{\mathrm{GC.US}} \mathrm{GC.US}_i \\
& +\beta_{\mathrm{GC.AP}} \mathrm{GC.AP}_i  +\beta_{\mathrm{GC.EU}} \mathrm{GC.EU}_i +\beta_{\mathrm{GC.UK}} \mathrm{GC.UK}_i \\
& +\beta_{\mathrm{UU}} \mathrm{US}_i \times \mathrm{GC.US}_i+\beta_{\mathrm{JA}} \mathrm{JP}_i \times \mathrm{GC.AP}_i+\beta_{\mathrm{EE}} \mathrm{EU}_i \times \mathrm{GC.EU}_i+\epsilon_i.\\
\end{aligned}\label{eq:linear_regression_full}
\end{equation}
Here, $S_i^{\mathrm{CAT}}$ denotes the spread in basis points for the $i$th bond. The model includes an intercept term, $\alpha$, 19 predictors, and 3 interaction terms. $\mathrm{EL}_i$, $\mathrm{PFL}_i$, $\mathrm{SIZE}_i$, $\mathrm{TERM}_i$, and $\mathrm{INDEM}_i$ represent expected loss (in percentage points), probability of first loss (in percentage points), issue volume (in USD million), term (in months), and trigger type (indemnity trigger or not) for CAT bond $i$, respectively.

Our model incorporates various dummy variables, each of which takes a value of 1 when a specific condition is met, and 0 otherwise. The dummy variable $\mathrm{INDEM}_i$ is set to 1 when CAT bond $i$ uses an indemnity trigger. Similarly, $\mathrm{WIND}_i$ and $\mathrm{EQ}_i$ are assigned a value of 1 for bonds referencing windstorms (e.g., hurricanes or tornadoes) and earthquakes, respectively. When both $\mathrm{WIND}_i$ and $\mathrm{EQ}_i$ are assigned a value of 0, it indicates that the bond covers either a different peril or multiple perils. Thus, we combine multiperil and other perils into a single category, which we then use as the reference level.
The dummy variables for covered territories include  $\mathrm{US}_i$, $\mathrm{EU}_i$, and $\mathrm{JP}_i$, where the value of 1 indicates the connection of a bond to a specific territory (e.g., US or Japan). If none of these variables is set to 1, the bond is tied to a different territory (e.g., Mexico) or multiple territories. Furthermore, $\mathrm{SR}_i$ indicates whether CAT bond $i$ is sponsored by Swiss Re, the largest CAT bond sponsor so far, and $\mathrm{IG}_i$ is set to one for investment-grade bonds.

The model also includes seven variables to capture market-wide cyclical spread drivers. $\mathrm{ROLX}_i$ represents the Lane Financial LLC Synthetic RoL Index, and $\mathrm{BBSPR}_i$ denotes the yield spread of BB-rated bonds. The five variables, $\mathrm{GC.GLOB}_i$, $\mathrm{GC.US}_i$, $\mathrm{GC.AP}_i$, $\mathrm{GC.EU}_i$, and $\mathrm{GC.UK}_i$, correspond to the Guy Carpenter Global Property Catastrophe RoL index for the global, US, Asia Pacific, Europe, and UK regions, respectively. For the BB spread, the average value for the quarter before bond issuance is used, while for the remaining indices, the value from the previous year (or quarter) of bond issuance is used. Interaction terms are considered for the most common combinations of territory and Guy Carpenter RoL indices, including US$\times$GC.US, JP$\times$GC.AP, and EU$\times$GC.EU.

To address the potential overfitting problem, we employ variable selection techniques, including LASSO and forward/backward stepwise selection. Table~\ref{tab:LR} presents the estimation results for the full model \eqref{eq:linear_regression_full}, LASSO regression, and the optimal model (with the smallest AIC value) resulting from the stepwise selection. All models are applied to the full sample of 765 bonds.

In the full model, the $p$-value of the coefficient for EL is close to zero, indicating its strong explanatory power for CAT bond spreads. The coefficient of EL is 179, meaning that a 1\% increase in EL would lead to a 1.79\% increment of the bond spread, all else being equal. The strong correlation between EL and bond spread is consistent with findings in the literature. Nevertheless, approximately one-third of the parameters are statistically insignificant, indicating the need for variable selection.  

We also find that the coefficient estimate for PFL is negative and statistically insignificant, which appears counter-intuitive. A positive correlation between PFL and CAT bond spreads is well-documented within the literature, as evidenced by \cite{lane2000pricing}, \cite{wills2010securitization}, and \cite{beer2022market}. This result  can be attributed to a high degree of collinearity between PFL and EL, as indicated by Figures~\ref{fig:spread_EL_sample} and \ref{fig:spread_PFL_sample}. Such collinearity can lead to unstable coefficient estimates and inflated standard errors. While collinearity may not compromise the overall explanatory capacity of the model, it poses challenges for the interpretability of the regression coefficients. For the same reason, coefficients for the Guy Carpenter indices and their $p$-values should be interpreted with caution due to the multi-collinearity among them.

The model derived from LASSO regression removes six variables -- PFL, WIND, US, EU, GC.UK, and JP$\times$GC.AP -- from the full model, resulting in slightly lower $R^2$ and adjusted $R^2$ compared to the full model. The optimal model from step-wise variable selection excludes PFL, SIZE, INDEM, WIND, US, EU, and JP, six of which were found to be insignificant in the full model. By excluding these non-significant variables, the optimal step-wise model achieves the highest adjusted $R^2$ among the three models assessed. Moreover, all variables included in the optimal model are statistically significant. 

The coefficients in the selected model exhibit the expected signs. For example, bonds covering earthquakes have on average lower spreads, as do bonds issued by Swiss Re or those with an investment-grade rating. It is worth noting that the negative coefficient associated with SR does not contradict the descriptive statistics in Table~\ref{Different Categorie}, where bonds issued by Swiss Re are shown to have higher spreads on average. This discrepancy arises because bonds issued by Swiss Re typically have a higher average EL, leading to higher spreads. However, when controlling for other factors, the linear regression model indicates that having Swiss Re as the issuer tends to lower bond spreads. 

\begin{table}[htbp]
\begin{adjustbox}{angle=90}

\begin{minipage}{\textheight}
\centering

\caption{Estimation results of the linear regression model, including the full model \eqref{eq:linear_regression_full}, LASSO, and the optimal model selected using step-wise backward and forward variable selection, respectively. Significance levels:  ‘***’ $<0.001$, ‘**’ $<0.01$, ‘*’ $<0.05$, ‘.’ $<0.1$.}
\begin{tabular}{llllllllllll}\toprule
              & \multicolumn{4}{c}{Full Model}    &  & LASSO    &  & \multicolumn{4}{c}{Step-wise Selection} \\ \midrule
              & Coeff.   & Std.   & p-Val. & Sig. &  & Coeff.   &  & Coeff.     & Std.    & p-Val.  & Sig.  \\
(Intercept)     & -149.58 & 120.39 & 0.21            &     &  & -177.64 &  & -174.13 & 104.46 & 0.10            & .   \\
EL              & 179.05  & 17.22  & \textless 2e-16 & *** &  & 165.41  &  & 165.68  & 4.23   & \textless 2e-16 & *** \\
PFL             & -9.03   & 12.40  & 0.47            &     &  & .       &  &         &        &                 &     \\
SIZE            & 0.11    & 0.09   & 0.21            &     &  & 0.05    &  &         &        &                 &     \\
TERM            & -3.34   & 0.93   & 0.00            & *** &  & -3.30   &  & -3.12   & 0.88   & 0.00            & *** \\
INDEM           & 14.62   & 23.15  & 0.53            &     &  & 1.19    &  &         &        &                 &     \\
WIND            & -1.64   & 28.51  & 0.95            &     &  & .       &  &         &        &                 &     \\
EQ              & -182.26 & 31.86  & 0.00            & *** &  & -171.87 &  & -176.69 & 28.16  & 0.00            & *** \\
US              & -61.24  & 90.76  & 0.50            &     &  & .       &  &         &        &                 &     \\
EU              & 20.64   & 198.08 & 0.92            &     &  & .       &  &         &        &                 &     \\
JP              & -516.14 & 270.93 & 0.06            & .   &  & -42.72  &  &         &        &                 &     \\
SR              & -65.96  & 23.34  & 0.00            & **  &  & -58.17  &  & -68.53  & 22.91  & 0.00            & **  \\
IG              & -200.22 & 53.18  & 0.00            & *** &  & -187.51 &  & -202.21 & 52.81  & 0.00            & *** \\
ROLX            & 3.22    & 0.83   & 0.00            & *** &  & 3.34    &  & 3.23    & 0.82   & 0.00            & *** \\
BBSPR           & 0.32    & 0.08   & 0.00            & *** &  & 0.26    &  & 0.32    & 0.08   & 0.00            & *** \\
GC.GLOB         & 2.41    & 0.68   & 0.00            & *** &  & 2.37    &  & 2.39    & 0.67   & 0.00            & *** \\
GC.US           & -2.14   & 0.50   & 0.00            & *** &  & -2.15   &  & -1.89   & 0.42   & 0.00            & *** \\
GC.AP           & -2.78   & 0.59   & 0.00            & *** &  & -1.76   &  & -2.65   & 0.58   & 0.00            & *** \\
GC.EU           & 7.84    & 2.76   & 0.00            & **  &  & 2.86    &  & 7.75    & 2.72   & 0.00            & **  \\
GC.UK           & -7.08   & 4.10   & 0.08            & .   &  & .       &  & -7.26   & 4.06   & 0.07            & .   \\
JP$\times$GC.AP & 2.70    & 1.52   & 0.08            & .   &  & .       &  & 2.76    & 1.51   & 0.07            & .   \\
US$\times$GC.US & 0.83    & 0.50   & 0.10            & .   &  & 0.46    &  & 0.51    & 0.12   & 0.00            & *** \\
EU$\times$GC.EU & -1.73   & 1.50   & 0.25            &     &  & -1.51   &  & -1.59   & 0.30   & 0.00            & *** \\
                &         &        &                 &     &  &         &  &         &        &                 &     \\
$R$-squared       & 0.775   &        &                 &     &  & 0.772   &  & 0.7741  &        &                 &     \\
$R$-squared Adj   & 0.7684  &        &                 &     &  & 0.7635  &  & 0.7693  &        &                 &   \\ \bottomrule
\end{tabular}\label{tab:LR}
\end{minipage}
\end{adjustbox}
\end{table}

\section{XGBoost: Model Estimation and Implications} \label{sec:rf}

We now proceed to the application of XGBoost models to CAT bond pricing. In these models, we incorporate all the variables in Equation \eqref{eq:linear_regression_full}, except interaction terms. The exclusion of interaction terms is deliberate, as machine learning models inherently capture interactions between variables. We keep both PFL and EL in the fitting of XGBoost models since tree-based methods are less sensitive to collinearity. However, high collinearity can still affect the interpretation of tree-based models, particularly when it comes to understanding the importance or impact of individual predictor variables. We will discuss this further in Section~\ref{sec:aleplot}. To avoid overfitting and improve model performance, machine learning methods often require careful tuning of hyperparameters.  This tuning process employs an exhaustive grid search of a predefined set of hyper-parameters to identify the combination that optimizes cross-validation prediction accuracy. In our approach, a 5-fold cross-validation was used during the grid search to ensure robust hyperparameter selection. After identifying the optimal hyperparameters, the model was then trained on the entire dataset. Key hyperparameters for both models and their optimal values obtained through grid search are presented in Table~\ref{RFhyperparameters}.

In the subsequent sections, we discuss the estimated XGBoost model and the implied impacts of various risk factors on CAT bond spreads. To achieve this, we employ feature importance and accumulated local effects (ALE) plots, both widely used tools for interpreting machine-learning-based models.

\subsection{Relative Importance of Risk Factors}

\begin{figure}[h!]
\begin{center}
\hspace*{-5cm}
			\includegraphics[width=0.75\linewidth]{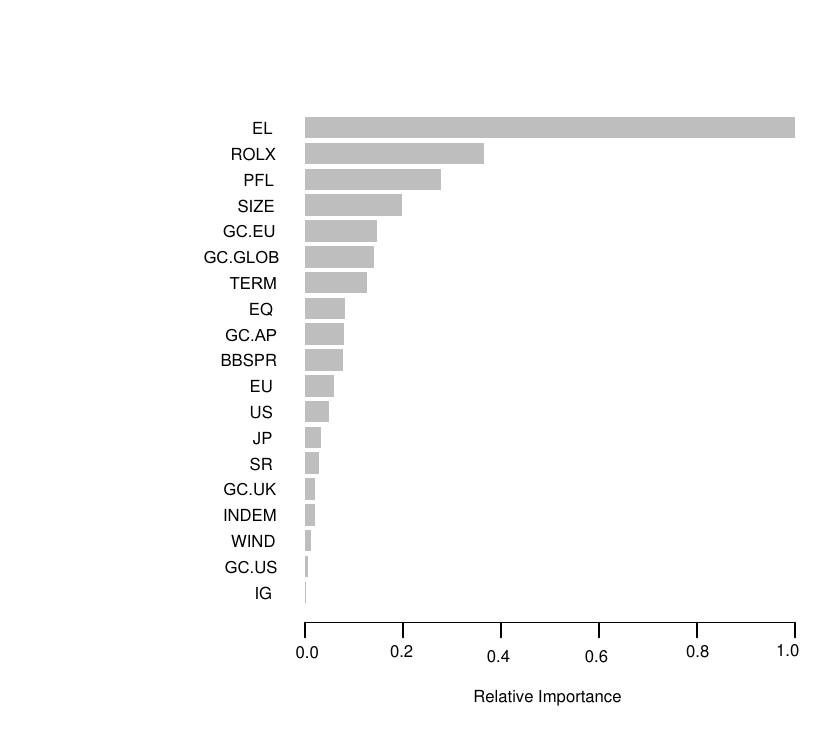}
\end{center}
	\caption{Feature importance plot for the XGBoost model.}
	\label{fig:var_importance_xgb}
\end{figure}

Figure~\ref{fig:var_importance_xgb} presents the feature importance plot which illustrates the relative importance of each risk factor in the trained XGBoost model, measured by its contribution to the model's goodness of fit. The plot shows that expected loss is the most important variable in explaining CAT bond spreads, consistent with findings from existing studies and the linear regression models. Notably, unlike the linear regression models, PFL also plays an important role in improving the explanatory power of the XGBoost model, even when EL is included.  This suggests that despite the high correlation between EL and PFL, the latter still provides additional valuable information for understanding CAT bond spreads. The linear regression model likely excludes PFL due to its nonlinear relationship with spreads, a characteristic that XGBoost is well-equipped to capture.

Beyond EL and PFL, the roles of other variables in the model are also noteworthy.  The Lane Financial LLC Synthetic RoL Index, $\mathrm{ROLX}$, ranks as the second most important variable in the XGBoost model. 
Additionally, the Guy Carpenter indices, $\mathrm{GC.EU}$ and $\mathrm{GC.GLOB}$, which track changes in reinsurance pricing for property catastrophe coverage in the European and global markets respectively, rank as the fifth and sixth most important features. These results align with the fact that CAT bond investors expect returns that reflect the risk they are assuming, often benchmarking these returns against the cost of traditional reinsurance. When reinsurance rates rise, CAT bond yields tend to increase as well. Moreover, reinsurance pricing reflects market perception of catastrophe risk. Changes in these prices can signal shifts in perceived risk, thereby influencing CAT bond pricing as investors and issuers reassess the risk-return profile. The importance of the three indices -- ROLX, GC.EU, and GC.GLOB -- is also evident in the linear regression model.

$\mathrm{SIZE}$ ranks as the fourth most important variable in the XGBoost model, despite being excluded from the stepwise-selected linear regression model (c.f. Table~\ref{tab:LR}). The exclusion of $\mathrm{SIZE}$ from the linear regression model may stem from its nonlinear relationship with bond spreads and potential collinearity with other variables, which complicates isolating its unique contribution within a linear framework. The XGBoost model, however, can capture nonlinear relationships and interactions and is more robust to collinearity, as it selects features based on their contribution to improving performance across multiple decision trees. Consequently, the XGBoost model recognizes the importance of $\mathrm{SIZE}$ even when correlated risk factors are present.
Finally, variables considered significant in the linear regression model, such as $\mathrm{IG}$ and $\mathrm{GC.US}$, contribute minimally to the goodness of fit in the XGBoost model.

\subsection{Impacts of Individual Risk Factors} \label{sec:aleplot}

\begin{figure}[h!]
\begin{center}
			\includegraphics[width=\linewidth]{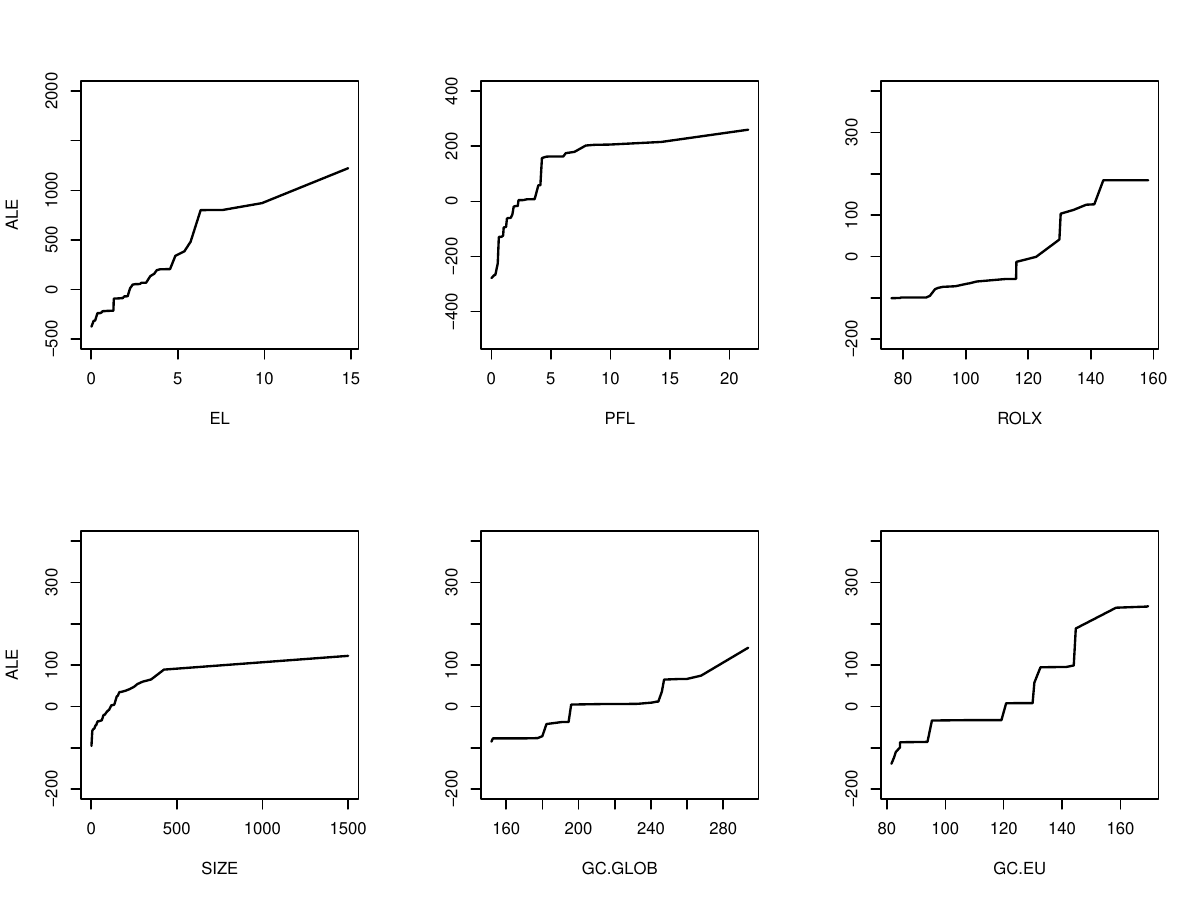}
\end{center}
	\caption{Accumulated local effects plots for the six most important predictors in the XGBoost model.}
	\label{fig:pdp_xgb}
\end{figure}

The feature importance plot provides an initial overview of each risk factor's relative contribution. However, this plot alone does not reveal how each factor influences CAT bond spreads. To address this, we further examine the impact of the six most important factors on CAT bond spreads by analyzing their accumulated local effect (ALE) plots, as shown in Figure~\ref{fig:pdp_xgb}. All ALE plots are centered to have an average of zero for enhanced interpretability. ALE plots are similar to the widely used partial dependence plots (PDP) but correct for the bias that arises in PDP when explanatory variables are highly correlated \citep{apley2020visualizing}. For instance, since EL and PFL are highly correlated, it is unlikely that a CAT bond would simultaneously have a very high EL and very low PFL, or vice versa. PDPs cannot adequately address such situations because they rely on the marginal distributions of the explanatory variables. Consequently, when evaluating the impact of EL on bond spreads, PDPs treat EL independently of other variables and may consider unrealistic cases where EL is high and PFL is very low. In contrast, ALE plots rely on conditional distributions, accounting for the correlation among variables,  and leading to more accurate interpretations of the model. Therefore, ALE plots are a more appropriate choice for our analysis, considering the potential high correlations among risk factors.

From Figure~\ref{fig:pdp_xgb}, nonlinear patterns are observed in the ALE plots. First, EL shows the strongest impact on bond spreads. When EL is close to zero, the bond spread is approximately 350 bps lower than the average bond spread. As EL increases to 15\%, the bond spread rises to approximately 1,250 bps above the average bond spread. For EL values below 5\%, there is an almost linear relationship, where bond spreads increase with the expected loss. Beyond the 5\% threshold, the relationship becomes less straightforward. A sharp increase in bond spread is observed as EL increases from 5\% to 6.25\%. However, as EL progresses from 6.25\% to 10\%, the increase in bond spread is minimal. For EL values above 10\%, bond spreads experience a modest increase with EL. Caution is advised when interpreting the section with large EL values exceeding 10\%, due to the limited number of observations in this range. 

Second, the ALE plot for PFL indicates that higher probabilities of first loss are associated with increased bond spreads, consistent with existing literature \citep{beer2022market}, which identifies a positive correlation between PFL and bond spread. However, our interpretation differs significantly. In \cite{beer2022market}, where only PFL is used as an explanatory variable, the effect of EL is indirectly captured by PFL due to their high correlation. Consequently, the impact attributed to PFL in their model is a combined effect of both PFL and EL.
In contrast, our XGBoost model incorporates both PFL and EL as explanatory variables, with EL being identified as the most significant factor. Given their strong correlation, much of the information represented by PFL is also encompassed within EL. As a result, PFL contributes to the XGBoost model with information that is not already accounted for by EL. Therefore, the XGBoost model effectively utilizes all the information contained in EL and PFL. The impact of this additional information in PFL is captured in the ALE plot. 

The impact of PFL on the bond spreads is most significant at lower levels of PFL and diminishes to almost zero as PFL exceeds 5\%. This indicates that at lower PFL values, investors are sensitive to changes in PFL and demand a higher spread for an increased PFL. However, as PFL reaches higher levels, further increases in PFL appear to only have a moderate impact on bond spread. A likely explanation for this observation is that when PFL is high, EL tends to be high as well due to their correlation. In such cases, investors may become less sensitive to small increases in PFL, shifting their focus more towards EL as the more influential factor in their decision-making process.

Third, the impacts of ROLX and GC.EU on bond spreads also exhibit clear non-linear patterns. The impact of GC.GLOB on the bond spread is similar to but less significant than that of ROLX and GC.EU.
Both ROLX and GC.EU have a modest influence on bond spread when their values are below 120, indicating a relatively soft reinsurance market.  In a soft market, there is generally an abundance of capital and a competitive environment. Investors might also perceive lower risk in this environment, resulting in a lower impact of changes in reinsurance market conditions on CAT bond spreads. Additionally, CAT bonds can be viewed as alternatives to traditional reinsurance products. Reinsurance remains the dominant risk transfer mechanism for catastrophe risks, and the choice between reinsurance and CAT bonds reflects a trade-off between their relative benefits and costs \citep{SUBRAMANIAN2018}. In a soft reinsurance market, where reinsurance capacity is ample and premiums are low, entities with hedging needs are more inclined to opt for reinsurance products rather than issuing CAT bonds. Small increases in reinsurance premiums are unlikely to push these entities toward the CAT bond market, as the increased premiums remain affordable. Therefore, CAT bond spreads remain low and relatively insensitive to reinsurance market conditions when the market is soft. 

As the reinsurance market hardens, indicated by increased ROLX and GC indices, capital becomes tighter, demand for coverage rises, and the perception of risk intensifies. Investors seek higher returns for assuming the same risks, leading to a more pronounced impact of ROLX and GC.EU on bond spreads, as shown by the steep increase in their corresponding ALE plots.  Due to limited capacity and high premiums in the reinsurance market, more entities with hedging demands turn to CAT bonds to transfer their risks. If reinsurance premiums continue to rise, more entities may shift towards the CAT bond market, driving up CAT bond spreads. This explains why CAT bond spreads are more sensitive to reinsurance market conditions when the market is hard.

Our analysis also indicates that CAT bond spreads tend to increase with issue size, particularly for issues smaller than 500 million US dollars. This observation contrasts the findings of \cite{braun2016pricing}, who reported a negative impact of issue size on spreads. This difference results from the incorporation of GC indices in our model, which exhibit a moderate negative correlation (Pearson correlation of $-0.27$ for GC.EU and $-0.21$ for GC.GLOB) with the issue size. The negative correlation suggests that in a hardening market, characterized by tighter capital and higher perceived risks, CAT bond issue sizes tend to decrease.
Therefore, issue size may serve as a weak indicator of market cycle conditions. In the absence of GC indices as predictors, as in the analysis by \cite{braun2016pricing}, the observed negative impact of issue size on spreads could primarily reflect the effect of a hardening market. By incorporating GC indices as risk factors in our model, we can distinguish the specific impact of issue size on bond spreads from the effects captured by the GC indices. The ALE plot suggests that in a hard market, where issue size is often small, an increase in issue size is associated with an increase in bond spreads. 

Overall, the ALE plots highlight significant nonlinear relationships between key risk factors and CAT bond spreads. Failing to account for these nonlinear patterns could potentially lead to inaccurate estimations of bond spreads.

\subsection{Interaction Effects} \label{sec:2nd_ale}

Machine learning models like Random Forest and XGBoost naturally account for interaction effects between explanatory variables, providing an advantage over linear regression, where interactions must be explicitly defined. To visualize these interactions within the estimated XGBoost model, we utilize second-order ALE plots. While first-order ALE plots illustrate the main effect of individual risk factors, second-order ALE plots reveal the \textit{additional} joint impact of pairs of risk factors on CAT bond spreads.

\begin{figure}
    \centering
    \begin{subfigure}[b]{0.495\textwidth}
    \includegraphics[width=\linewidth]{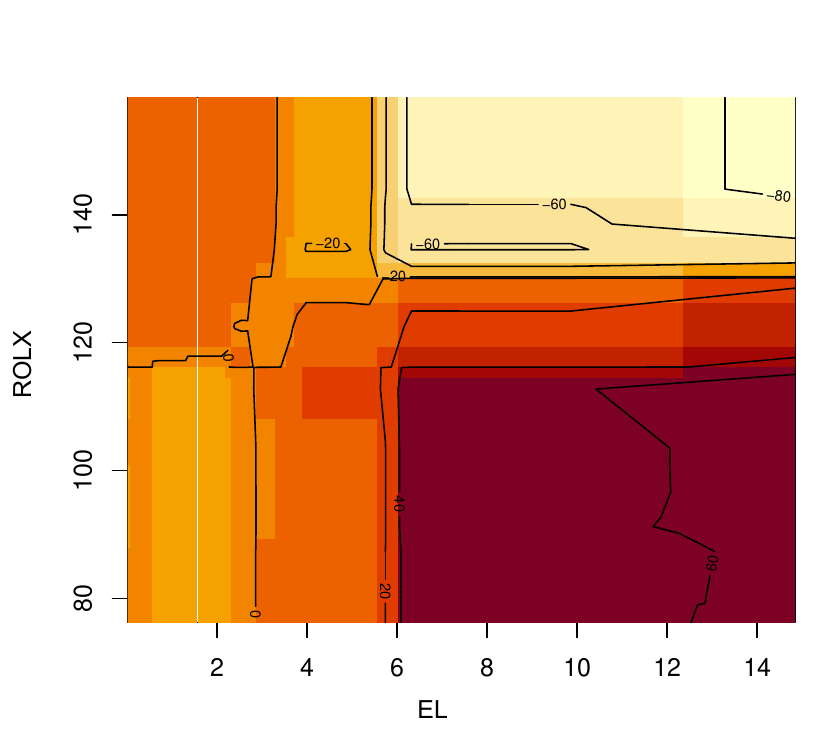}
    \caption{EL (x-axis) and ROLX (y-axis)}
    \label{fig:interact_ELROLX}
    \end{subfigure}
\begin{subfigure}[b]{0.495\textwidth}
    \includegraphics[width=\linewidth]{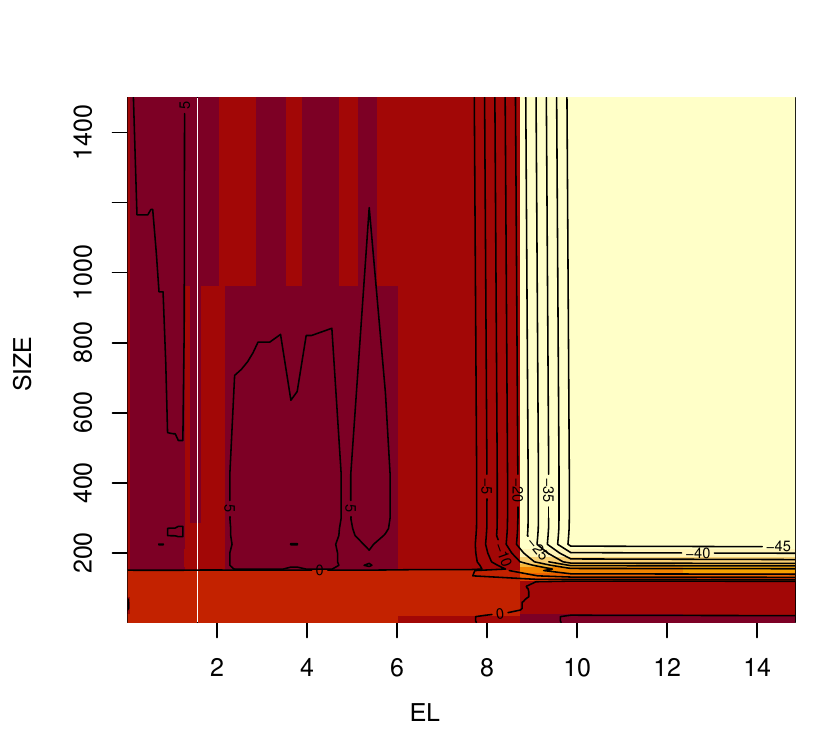}
    \caption{EL (x-axis) and SIZE (y-axis)}
    \label{fig:interact_ELSIZE}
\end{subfigure}
\begin{subfigure}[b]{0.495\textwidth}
    \includegraphics[width=\linewidth]{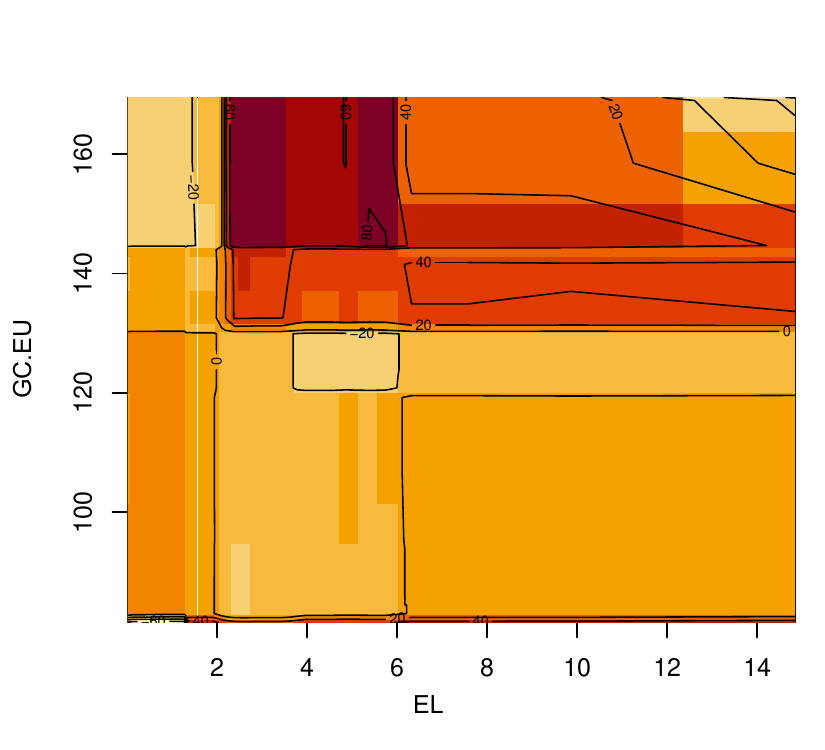}
    \caption{EL (x-axis) and GC.EU (y-axis)}
    \label{fig:interact_ELEU}
\end{subfigure}
\begin{subfigure}[b]{0.495\textwidth}
    \includegraphics[width=\linewidth]{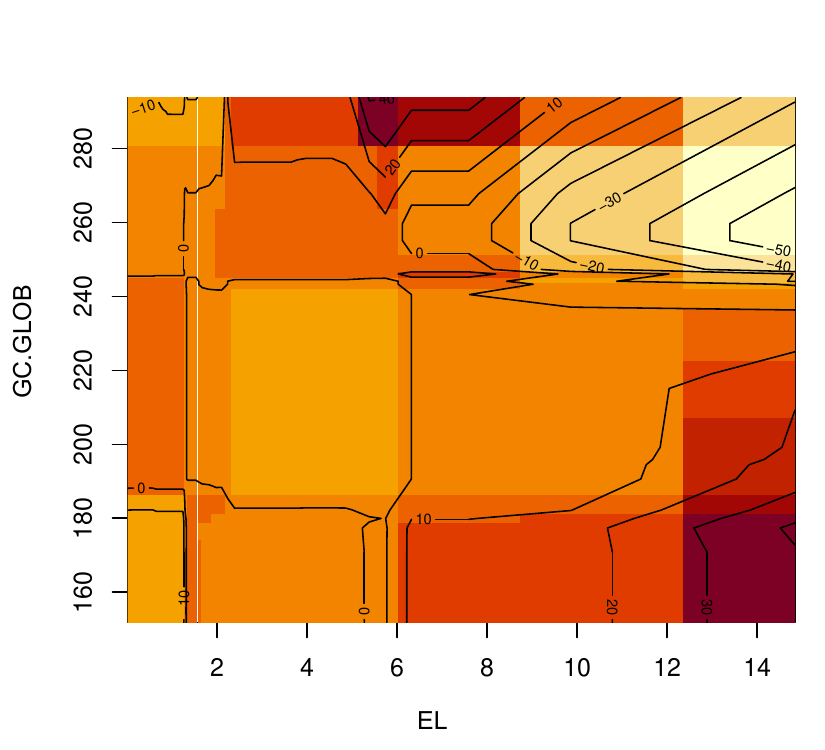}
    \caption{EL (x-axis) and GC.GLOB (y-axis)}
    \label{fig:interact_ELGLOB}
\end{subfigure}
   \caption{Second-order ALE plots of four selected predictor pairs.}
   \label{fig:ALE_2order}
\end{figure}

As an illustration, Figure \ref{fig:ALE_2order} displays second-order ALE plots for four pairs of risk factors: EL \& ROLX, EL \& SIZE, EL \& GC.EU, and EL \& GC.GLOB. The contour values on these plots indicate the strength of the joint effects. All four plots demonstrate clear interaction effects between the respective pairs of risk factors, as  the contour values vary significantly in each plot. The magnitude of these values is smaller compared to the ALE values depicted in Figure \ref{fig:pdp_xgb}, suggesting that while these second-order effects are significant, they are generally smaller than the main effects of the individual risk factors.

We first examine Figure \ref{fig:interact_ELROLX}, which displays the interaction between EL and ROLX. At an EL value of 2\%, an increase in ROLX from 120 to 140 does not change the spread. In contrast, at an EL of 8\%, the same increase in ROLX results in an 80 bps decrease in the spread, thereby suggesting a negative interaction between EL and ROLX. It is important to note that the second-order ALE effect is distinct from the main effect. To fully understand the impact of ROLX on the spread given the EL value, one must also consider the main effect of an increase in ROLX, which is positive within the range of 120--140. Specifically, an increase in ROLX from 120 to 140 results in approximately a 140 bps increase in the spread of a bond with a 2\% EL, as illustrated in Figure~\ref{fig:pdp_xgb}. However, this increase translates to only a 60 bps (140 bps $-$ 80 bps) spread increase when EL is 8\%. Therefore, for bonds with a higher EL, the impact of ROLX on the spread remains positive but weakened after combining the main and second-order effects.

Analysis of Figure \ref{fig:interact_ELSIZE} indicates that at an EL value of 2\%, an increase in SIZE from 150 to 250 has a minimal effect on the spread. However, the same increase in SIZE results in approximately 45 bps decrease in the spread when EL is 10\%, suggesting a negative interaction between EL and SIZE. We observed a positive main effect of SIZE on bond spread in Figure \ref{fig:pdp_xgb}. The negative interaction between EL and SIZE suggests that an increase in issue size leads to a less pronounced rise, or even a decrease, in the spread for bonds with a higher EL. Given that the main effect of SIZE on the spread is relatively modest, as depicted in Figure \ref{fig:pdp_xgb}, it is conceivable that the combined main and interaction effects of certain increases in SIZE could result in a negative impact on the spread for bonds with a high EL.

Figures \ref{fig:interact_ELEU} and \ref{fig:interact_ELGLOB} depict the interactions between EL and the GC indices, GC.EU and GC.GLOB, respectively. These interactions exhibit a complex pattern, neither consistently positive nor negative. For instance, at an EL value of 2\%, increasing GC.EU from 140 to 160 does not affect the spread. At an EL of 4\%, this same increase in GC.EU is associated with an increase in the spread. Interestingly, at an EL of 8\%, an identical change in GC.EU results in a decrease in spread.  The XGBoost model, with its tree-based structure, can capture these complex interactions, offering greater flexibility and higher explanatory power than models that assume linear relationships.

\section{Prediction Performance and Conditional Impact}
\label{sec:prob_analysis}

In this section, we examine the prediction performance of linear regression models and XGBoost models by analyzing their prediction accuracy and prediction intervals. In addition, we explore the conditional impacts of various factors by observing how changes in one factor influence the predictions and prediction intervals while fixing other factors. The prediction intervals are produced from the normally distributed error terms for the linear regression model and from employing the \textit{Jackknife+} prediction intervals introduced in Section~\ref{sec:jackknife} for XGBoost. Section \ref{sec:PIs} compares the prediction accuracy and prediction intervals generated by these two approaches. Section \ref{sec:case_study} further investigates the conditional impact of expected loss and its interactions with other factors.

\subsection{Predictive Accuracy and Intervals}\label{sec:PIs}

In statistical learning literature, it is common practice to randomly split the data into two parts: a training set used to train the model and a test set used to evaluate its predictive performance. However, relying on a single split can introduce randomness, making the analysis less robust. To address this, we perform a Monte Carlo cross validation involving 1,000 random data splits.
For each split, 80\% of the observations are randomly selected as the training set. This training set is first used for hyperparameter tuning through cross-validation, followed by training the final model using the identified optimal parameters. The remaining 20\% of the data, which forms the test set, is then used strictly for prediction, allowing us to assess predictive performance. The mean squared errors (MSE), mean prediction interval length, and mean prediction interval coverage for the test set are calculated for each model and each split.  

The resulting average test MSEs across the 1,000 splits for the XGBoost and the linear regression models selected by the step-wise procedure and LASSO are presented in Table~\ref{tab:my_label}. For comparison purposes, we also include the test MSE for Random Forest. The two linear regression models produce similar forecasting results, while the two machine learning models demonstrate greater accuracy, generating lower MSEs than their linear counterparts.  XGBoost produces the lowest average test MSE across all models. 

\begin{table}[htbp]
    \centering
    \caption{Average test MSE using 1,000 random training/test splits.}
    \begin{tabular}{ll}\hline
        Model & Average test MSE\\\hline
        XGBoost & 0.0004580 \\
        Random Forest & 0.0005287 \\
        Linear regression (step-wise selection) &  0.0006487\\ 
        Linear regression with Lasso regularization &  0.0006439 \\ \bottomrule
        \end{tabular} 
    \label{tab:my_label}
\end{table}

Since the linear regression models selected using the step-wise method and Lasso regularization lead to similar predictions, we only consider the model selected by step-wise method in the following analysis. 
The average coverage and length of the 95\% prediction intervals generated by the linear regression and XGBoost with \textit{Jackknife+} based on the 1,000 random splits are presented in Table~\ref{tab:coverage}. The XGBoost model with \textit{Jackknife+} produces prediction intervals with an average coverage closer to the desired 95\% level and an average length 8\% smaller than those generated by the linear regression models. This suggests that the combination of Conformal Prediction with XGBoost yields more informative and precise prediction intervals.

\begin{table}[htbp]
    \centering
    \caption{Average coverage and length of the 95\% prediction intervals generated by the linear regression and XGBoost with \textit{Jackknife+} using 1,000 random training/test splits.}
\begin{tabular}{lll} \hline
Model                      & Average Coverage & Average Length \\ \hline
Linear Regression          & 0.9568           & 958.85          \\ \hline
XGBoost + Conformal Prediction & 0.9499           & 886.43       \\ \bottomrule
\end{tabular}
\label{tab:coverage}
\end{table}

\begin{figure}[htbp!]
\begin{center}
		\begin{subfigure}[b]{0.5\textwidth}
			\includegraphics[width=\linewidth]{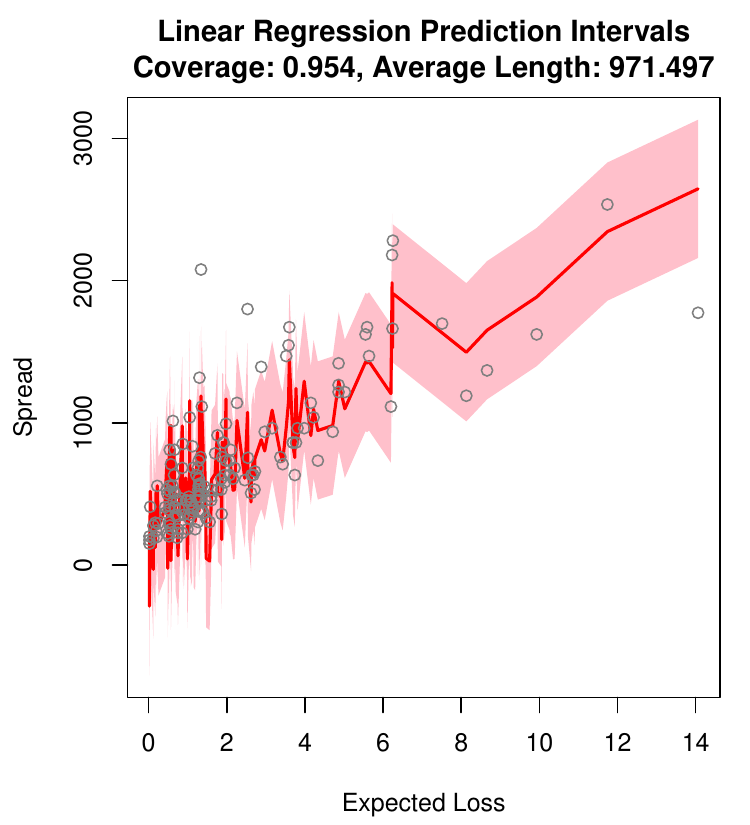}
	           \caption{Linear regression}
	           \label{fig:LR_CI}
            \end{subfigure}%
		\begin{subfigure}[b]{0.5\textwidth}
			\includegraphics[width=\linewidth]{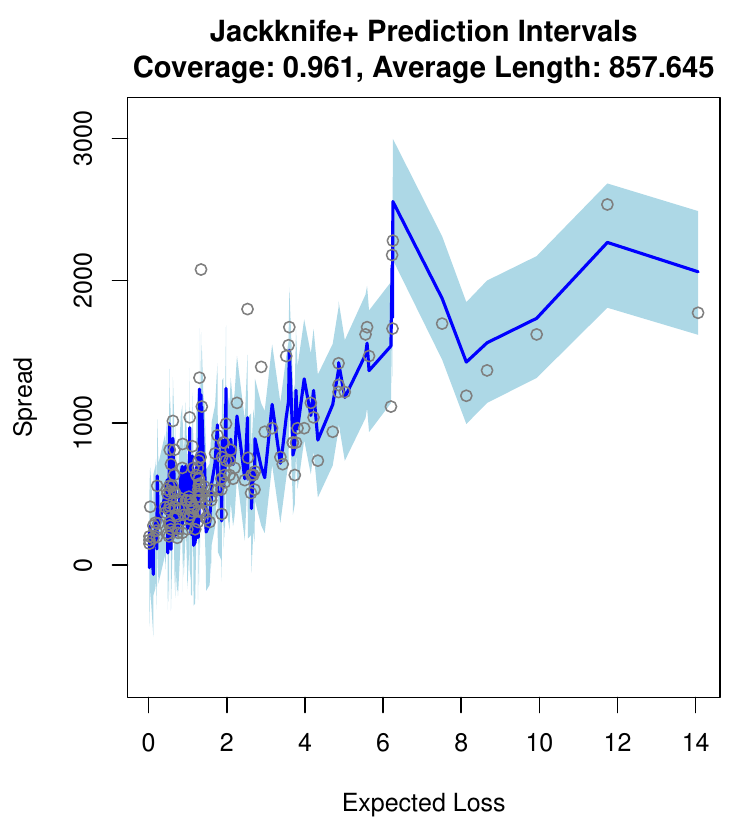}
	           \caption{XGBoost + Conformal Prediction}
	           \label{fig:XGB_CI}
		\end{subfigure}%
\end{center}
\caption{Mean forecasts and the 95\% prediction intervals for the test data generated by the linear regression and XGBoost with \textit{Jackknife+}.}
\label{fig:CI}
\end{figure}

Figure~\ref{fig:CI} presents the point forecasts (solid lines) and the corresponding 95\% prediction intervals (shaded areas) generated by both models for the test data based on a single split. The observed bond spreads (grey circles) are also depicted for reference and are arranged in order of their EL values.
We observe that both models produce prediction intervals with a coverage close to the 95\% target, with XGBoost achieving slightly higher coverage. Additionally, the average length of the prediction interval produced by XGBoost is approximately 13\% smaller than that generated by the linear regression model. This reaffirms our earlier finding that integrating Conformal Prediction with XGBoost leads to more precise and informative prediction intervals.

\subsection{Conditional Predictions}
\label{sec:case_study}

We further exmaine the conditional impact of several key risk factors using both linear regression and XGBoost models. Conditional impact refers to the effect of a single predictor on the predicted outcome, while keeping all other variables constant. To compute these conditional impacts, we initially calculate conditional predictions by varying the value of the risk factor under examination while holding all other factors at their median values for continuous variables or at the most common category for categorical variables. This process allows us to isolate and assess the impact of the specific factor. Additionally, prediction intervals can be computed for these pseudo-observations, offering insights into the variation in conditional predictions.

\begin{figure}[h!]
\begin{center}
			\includegraphics[width=0.6\linewidth]{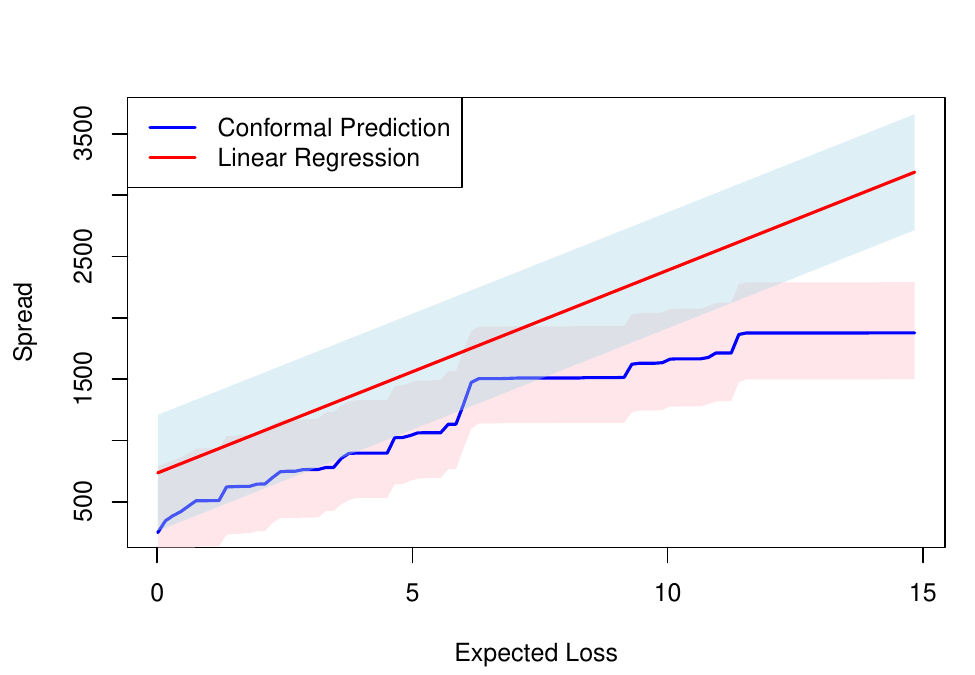}
\end{center}
	\caption{The conditional impact of expected loss on CAT bond spreads. All other risk factors are set to their median values (for continuous variables) or the most common type (for categorical variables).}
     \label{fig:case_study_EL}
\end{figure}

Figure~\ref{fig:case_study_EL} depicts the conditional predictions of bond spreads across various EL values generated by both the linear regression model (red line) and XGBoost (blue line), along with their corresponding 95\% prediction intervals (shaded areas). First, while the linear regression model predicts a linear impact of EL on CAT bond spreads, the XGBoost model exhibits nonlinear effects. Specifically, while increasing EL generally leads to larger spreads, this effect is more pronounced for smaller EL values and levels off when it exceeds 7\%. Second, the linear regression model consistently yields higher spreads compared to XGBoost for all EL values. This observation suggests that the linear regression model tends to overestimate bond spreads, particularly for larger EL values. Notably, the mean forecast produced by the linear regression model lies outside the 95\% prediction interval generated by Conformal Prediction for the majority of EL values considered. This finding indicates that the overestimation by the linear model could be substantial.

\begin{figure}[htbp]
\begin{center}
		\begin{subfigure}[b]{0.5\textwidth}
			\includegraphics[width=\linewidth]{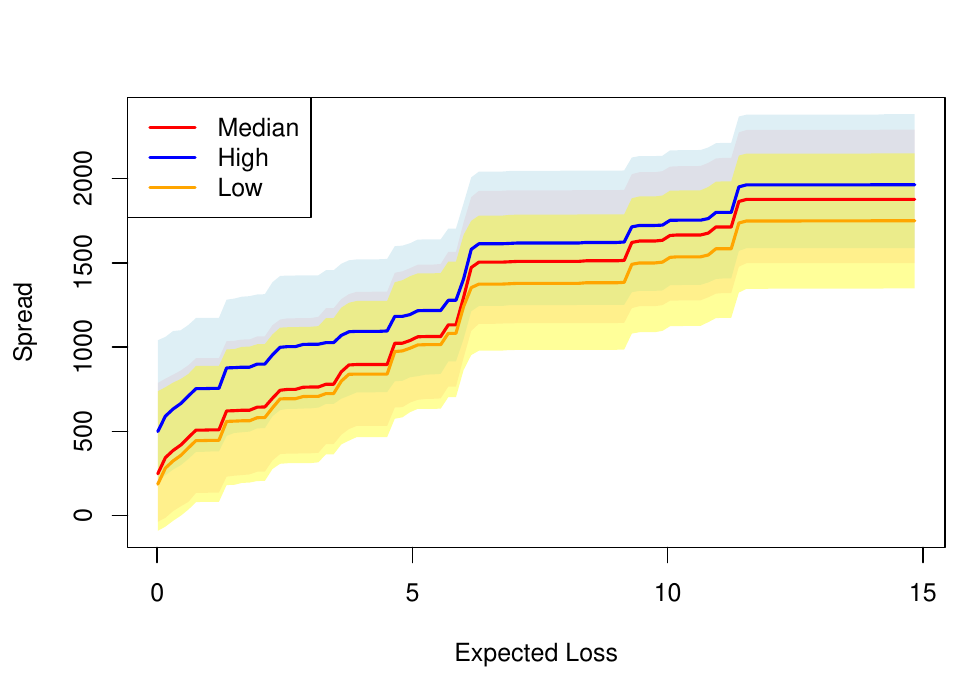}
	           \caption{XGBoost}
                \label{fig:rolx_xgb}
            \end{subfigure}%
		\begin{subfigure}[b]{0.5\textwidth}
			\includegraphics[width=\linewidth]{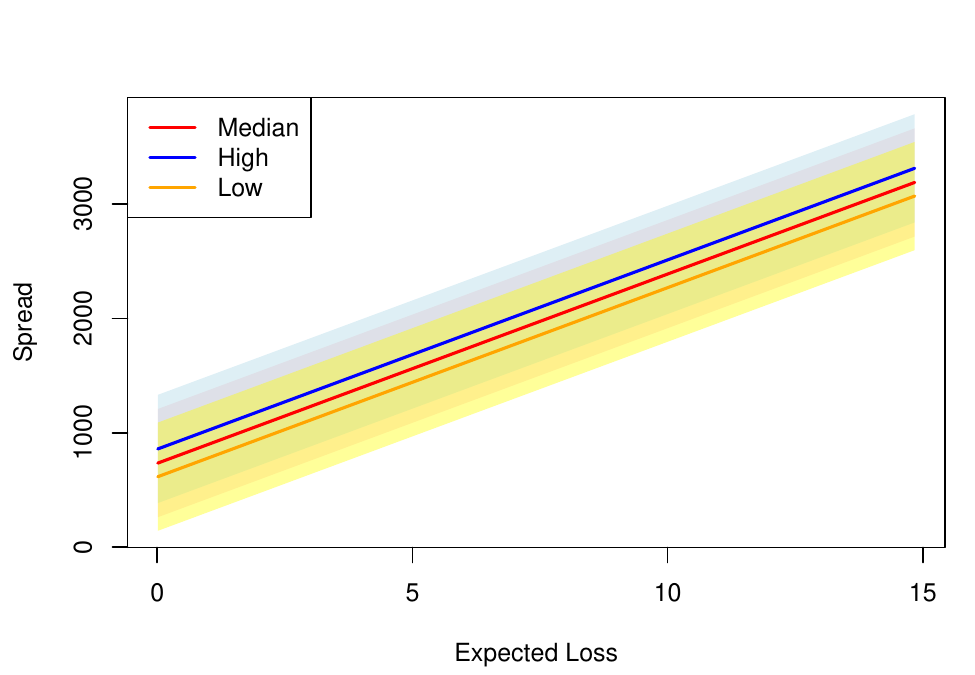}
	           \caption{Linear regression}
            \label{fig:rolx_lr}
		\end{subfigure}%
\end{center}
	\caption{Conditional impact of expected loss on CAT bond spreads under different reinsurance market conditions implied by XGBoost (left) and the linear regression model (right).}
 \label{fig:case_study_rolx}
\end{figure}

Figure~\ref{fig:case_study_EL} examines the conditional impact of EL on bond spread while all other risk factors are set at the median values or the most common categories. We now further investigate the impact of EL under different reinsurance market conditions, represented by ROLX values. Figure~\ref{fig:case_study_rolx} displays the conditional impact of EL across three reinsurance market conditions. The blue line represents a hard reinsurance market, where the ROLX value is set as the average of the top 10\% values, which is 144. The yellow line depicts a soft reinsurance market, where ROLX averages the bottom 10\% value, 88. Finally, the red line corresponds to a normal market, with ROLX set at the median value of 116. Figures~\ref{fig:rolx_xgb} and \ref{fig:rolx_lr} present the predictions generated by the XGBoost and linear regression models, respectively, along with 95\% prediction intervals for each conditional prediction.

Figure \ref{fig:rolx_xgb} demonstrates that with the XGBoost model, the shape of conditional prediction curves remains largely consistent across varying market conditions for different EL values, indicating a persistent nonlinear relationship between EL and spreads. However, the shifts in the conditional predictions across market conditions are not parallel. First, shifts from soft to normal market conditions result in only modest changes in conditional predictions for EL values below 7\%, as evidenced by the close alignment of the red and yellow lines over this interval. This observation suggests that transitions from soft to normal market conditions have minor effects on the spreads of CAT bonds with lower EL values. In contrast, a transition from normal to hard market conditions leads to a significant increase in conditional predictions for the same EL values. This indicates lower investor sensitivity to the reinsurance market cycle in softer conditions, whereas in hard markets, characterized by constrained capacity and rising premiums, investors seek higher returns to offset the risks associated with CAT bonds, even those with lower EL values. Second, the upward shifts of bond spreads are comparable in size for higher EL values when the reinsurance market transitions from soft to normal, and from normal to hard. This suggests that CAT bonds with higher EL values are sensitive to the tightening of reinsurance market conditions, even when the market is soft. These non-parallel changes in conditional impact curves across market conditions suggest an interaction between reinsurance market conditions and expected loss. In contrast, in Figure \ref{fig:rolx_lr}, where the linear regression model is applied, the conditional prediction curve shifts in parallel with changes in market conditions. The equal magnitude of shift when market conditions change reflects that the linear regression model only captures the linear effects of ROLX.


\begin{figure}[h!]
\begin{center}
			\includegraphics[width=0.6\linewidth]{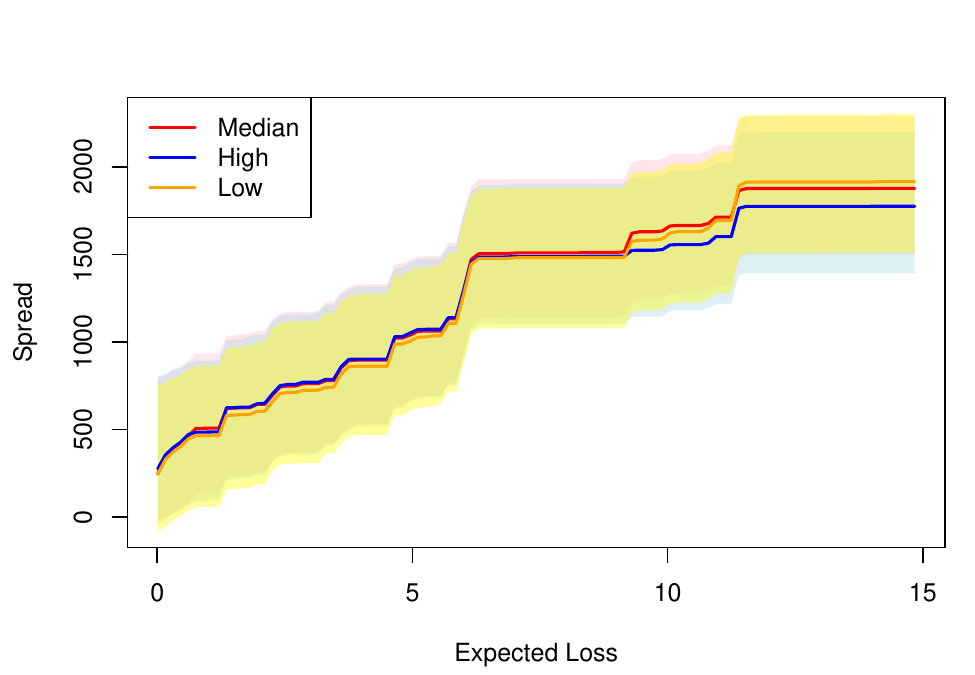}
\end{center}
	\caption{Conditional impact of expected loss on CAT bond spreads with different sizes.}
     \label{fig:case_study_ELsize}
\end{figure}

Figure~\ref{fig:case_study_ELsize} illustrates the conditional impact of EL for bonds with different issue sizes. The blue line represents a high issuance volume, where the SIZE value is set at the average of the top 10\% values, amounting to \$374.5 million. The yellow line depicts a low issuance volume, averaging the bottom 10\% value, which is \$14.96 million. The red line corresponds to a median SIZE value of \$100 million. Two observations, consistent with Figures~\ref{fig:pdp_xgb} and \ref{fig:interact_ELSIZE}, can be made. First, for small EL values, bonds with low issuance volumes generally have lower spreads compared to those with high or medium issuance volumes, indicating that SIZE has a positive impact on bond spreads. However, this impact appears to be relatively minor. Second, when EL is large, SIZE exhibits a negative impact on spreads. This negative impact is attributed to the interaction between SIZE and EL, which is stronger for high EL values, as discussed in Section~\ref{sec:2nd_ale}.

\section{Conclusion} \label{sec:conclusion}

In this study, we employ probabilistic machine learning techniques to analyze CAT bond pricing. Specifically, we use the XGBoost ensemble learning method to generate point estimates of bond spreads, accompanied by Conformal Prediction to construct prediction intervals. This machine learning method provides insights into the complex relationships between various risk factors and CAT bond spreads, as well as the interaction effects among these factors. It has also demonstrated improved accuracy and more reliable prediction intervals compared to widely used linear regression models. 


Our analysis utilizes a comprehensive dataset of primary market CAT bond transactions from January 1999 to March 2021. We begin with extensive linear regression models, followed by feature selection through Lasso Regression and a stepwise approach. This process identifies expected loss (EL) as a primary explanatory variable for CAT bond spreads, while another important factor, the probability of first loss (PFL), is excluded due to its strong correlation with EL. Interestingly, when employing the XGBoost model, PFL emerges as the third most important factor, indicating that despite its correlation with EL, PFL provides valuable information about CAT bond spreads not captured by EL alone. Nonlinear impacts on bond spreads have been observed from factors such as EL, PFL, issue size, and market-wide indices representing reinsurance market conditions. Furthermore, we uncover significant interaction effects among these factors. Finally, we find that the XGBoost model, combined with the Conformal Prediction method, consistently produces not only more accurate point forecasts but also prediction intervals that are closer to the desired coverage level and exhibit smaller lengths.

This study advances our understanding of CAT bond pricing dynamics through machine learning techniques. These advancements offer tangible benefits to stakeholders in the CAT bond market, including investors, issuers, and policymakers. The improved understanding of risk factors and market sensitivities we provide is crucial for effective risk management and strategic decision-making.
Future research could 
experiment with alternative machine learning algorithms, which may offer new insights into the evolving landscape of CAT bond pricing. Additionally, investigating the pricing dynamics of CAT bonds in the secondary market presents an intriguing avenue for future exploration.\\


\bibliography{reference}
\bibliographystyle{apalike}

\newpage
\begin{appendices}
    \section{GBM Algorithm}\label{apdx:gbm}
    A detailed GBM algorithm is provided below.
\begin{enumerate}
\item \textbf{Initialization}: Set $\hat{f}(\bm{x}_i)=c$ for all $i$ in the training set, where $c$ is a constant, typically chosen as the mean of the target variable in a regression setting.
\item \textbf{Residual Calculation}: Compute the residuals $r_i=y_i-\hat{f}(\bm{x}_i)$, which represent the differences between the observed and predicted values of the target variable.
\item \textbf{Fit Model to Residuals}: Fit a new weak model $\hat{f}_b$ to these residuals, indexed by the iteration number $b$. This objective is to predict and correct the errors made by the existing ensemble of models.
\item \textbf{Update Predictions}: Modify the prediction function $\hat{f}(\bm{x}_i)$ using the formula $\hat{f}(\bm{x}_i)\leftarrow \hat{f}(\bm{x}_i)+\eta\hat{f}_b(\bm{x}_i)$. Here, the predictions from the new week model $\hat{f}_b$ are scaled by a learning rate $\eta$ to mitigate overfitting, and then added to the previous predictions to form updated predictions.
\item \textbf{Update Residuals}: Adjust the residuals using $r_i\leftarrow r_i-\eta\hat{f}_b(\bm{x}_i)$. This update directs the algorithm to focus on the areas where the model is performing poorly in the current iteration.
\item \textbf{Iterate}: Repeat steps 2 to 5 for a specified number of iterations $B$. Each iteration refines the model by fitting a new weak learner to the updated residuals, progressively enhancing the prediction accuracy.
\item \textbf{Output Final Model}: The ultimate model is formulated as $\hat{f}(\bm{X})=\sum_{b=1}^{B}\eta\hat{f}_b(\bm{X})$, which represents an ensemble of scaled predictions from weak learners.
\end{enumerate}

\section{Glossary of Predictors and Optimal Hyperparameters} \label{sec:hyper_par}

Table~\ref{tbl:glossary} reports the predictors used in the empirical analysis, and Table~\ref{RFhyperparameters} reports the hyperparameter values of the Random Forest and XGBoost models tuned from the full sample of 765 bonds.

\begin{table}[htbp]
 \centering
  \begin{threeparttable}
    \caption{Glossary of predictors.}
     \begin{tabular}{p{1.8cm}p{11cm}p{1.7cm}}
\toprule
    Variable & Description &  Type  \\
\midrule
EL & Annual expected loss (in percentage points) & Continuous \\
PFL & Probability of first loss (in percentage  points) & Continuous\\
SIZE & Issue volume (in US million) & Continuous \\
TERM & CAT bond term (in months) & Continuous \\
INDEM & Dummy variable indicating whether the trigger type is indemnity or not) & Binary \\
WIND & Dummy variable indicating whether the CAT bond only covers losses from windstorms (e.g., hurricanes or tornadoes) & Binary \\
EQ & Dummy variable indicating whether the CAT bond only covers losses from earthquakes & Binary \\
US & Dummy variable indicating whether the covered territory is in US only & Binary  \\
EU & Dummy variable indicating whether the covered territory is in Europe only & Binary  \\
JP & Dummy variable indicating whether the covered territory is in Japan only& Binary  \\
SR & Dummy variable indicating whether the CAT bond is sponsored by Swiss Re & Binary \\
IG & Dummy variable indicating whether the bond is rated investment grade & Binary \\
ROLX & Quarterly values of Lane Financial LLC Synthetic Rate on Line Index & Continuous \\
BBSPR & Yield spread of BB-rated bonds over risk-free securities such as US treasuries bonds & Continuous \\
GC.GLOB & Guy Carpenter Global Property Catastrophe RoL index for the global property catastrophe insurance market & Continuous \\
GC.US &  Guy Carpenter Global Property Catastrophe RoL index for the US property catastrophe insurance market & Continuous  \\
GC.AP &   Guy Carpenter Global Property Catastrophe RoL index for the Asia-Pacific property catastrophe insurance market & Continuous \\
GC.EU & Guy Carpenter Global Property Catastrophe RoL index for the European property catastrophe insurance market &  Continuous \\
GC.UK &  Guy Carpenter Global Property Catastrophe RoL index for the global property catastrophe insurance the UK property catastrophe insurance market& Continuous \\
\bottomrule
    \end{tabular}
     \label{tbl:glossary}
  \end{threeparttable}
\end{table}

\begin{table}[htbp]
\caption{Optimal hyperparameters of theXGBoost models.}
\label{RFhyperparameters}
\begin{center}
\begin{tabular}{p{3cm}p{6cm}c} \toprule
\multicolumn{3}{c}{XGBoost} \\ \midrule
Hyperparameter & Description & Optimal Value \\  \midrule
\multirow{2}{*}{learning\_rate} & Learning rate that shrinks the contribution of each tree & \multirow{2}{*}{0.01} \\ 
{max\_depth} & Maximum depth of each tree & 4 \\ 
\multirow{2}{*}{min\_reduction} & Minimum loss reduction required to make a further partition on a leaf node of the tree & \multirow{2}{*}{0.00004} \\ 
\multirow{2}{*}{min\_child\_weight} & Minimum sum of instance weight needed in a leaf node & \multirow{2}{*}{5} \\ 
\multirow{2}{*}{subsample} & The ratio of subsample that XGBoost randomly collects to grow trees & \multirow{2}{*}{0.9} \\ 
{lambda} & $L2$ regularization term on weights & 400 \\ 
{nrounds} & Number of decision trees& 800 \\ 
\bottomrule
\end{tabular}
\end{center}
\end{table}

\end{appendices}

\end{document}